\documentclass{aa}
\usepackage{graphicx}
\usepackage[varg]{txfonts}
\begin{document}

\newcommand {\cm} {cm$^{-1}$}
\newcommand {\mc} {$\mu$m}
\newcommand {\wat} {H$_2$O}
\newcommand {\coo} {CO$_2$}
\newcommand {\meoh} {CH$_3$OH}
\newcommand {\met} {CH$_4$}
\newcommand {\oo} {O$_2$}
\newcommand {\nn} {N$_2$}
\newcommand {\nh} {NH$_3$}

   \authorrunning{Ioppolo et al.}
   \titlerunning{VUV spectroscopy of 1 keV electron irradiated N$_{2}$- and O$_{2}$-rich ices}
   \title{Vacuum ultraviolet photoabsorption spectroscopy of space-related ices: 1 keV electron irradiation of nitrogen- and oxygen-rich ices}

   \author{S. Ioppolo  \inst{1}
          \and
          Z.~Ka\v{n}uchov\'{a}  \inst{2}
          \and
          R.L. James  \inst{3}
          \and
          A. Dawes  \inst{3}
          \and
          N.C. Jones  \inst{4}
          \and
          S.V. Hoffmann  \inst{4}
          \and
          N.J. Mason  \inst{5}
          \and
          G. Strazzulla  \inst{6}
          }
\offprints{S. Ioppolo;  \email{s.ioppolo@qmul.ac.uk}}

\institute{School of Electronic Engineering and Computer Science, Queen Mary University of London, Mile End Road, London E1 4NS, UK\\
         \and
Astronomical Institute of Slovak Academy of Sciences, SK-059 60 Tatransk\'{a} Lomnica, Slovakia\\
         \and
School of Physical Sciences, The Open University, Walton Hall, Milton Keynes MK7 6AA, UK\\
         \and
ISA, Department of Physics and Astronomy, Aarhus University, Ny Munkegade 120, DK-8000 Aarhus C, Denmark\\
         \and
School of Physical Sciences, University of Kent, Park Wood Rd, Canterbury CT2 7NH, UK\\
         \and
INAF - Osservatorio Astrofisico di Catania, Via Santa Sofia 78, Catania I-95123, Italy\\
             }

    \date{Received ; accepted }

  \abstract
{Molecular oxygen, nitrogen, and ozone have been detected on some satellites of Saturn and Jupiter, as well as on comets. They are also expected to be present in ice-grain mantles within star-forming regions. The continuous energetic processing of icy objects in the Solar System induces physical and chemical changes within the ice. Laboratory experiments that simulate energetic processing (ions, photons, and electrons) of ices are therefore essential for interpreting and directing future astronomical observations.}
{We provide vacuum ultraviolet (VUV) photoabsorption spectroscopic data of energetically processed nitrogen- and oxygen-rich ices that will help to identify absorption bands and/or spectral slopes observed on icy objects in the Solar System and on ice-grain mantles of the interstellar medium.}
{We present VUV photoabsorption spectra of frozen  O$_{2}$ and N$_{2}$, a 1:1 mixture of both, and a new systematic set of pure and mixed nitrogen oxide ices. Spectra were obtained at 22~K  before and after 1~keV electron bombardment of the ice sample. Ices were then annealed to higher temperatures to study their thermal evolution. In addition, Fourier-transform infrared spectroscopy was used as a secondary probe of molecular synthesis to better identify the physical and chemical processes at play.}
{Our VUV data show that ozone and the azide radical (N$_3$) are observed in our experiments after electron irradiation of pure O$_{2}$ and N$_{2}$ ices, respectively. Energetic processing of an O$_{2}$:N$_{2}$~=~1:1 ice mixture leads to the formation of ozone along with a series of nitrogen oxides. The electron irradiation of solid nitrogen oxides, pure and in mixtures, induces the formation of new species such as O$_{2}$, N$_{2}$ , and other nitrogen oxides not present in the initial ice. Results are discussed here in light of their relevance to various astrophysical environments. Finally, we show that VUV spectra of solid NO$_{2}$ and water can reproduce the observational VUV profile of the cold surface of Enceladus, Dione, and Rhea, strongly suggesting the presence of nitrogen oxides on the surface of the icy Saturn moons.}
{}


   \maketitle

\section{Introduction}

It is well known that ices in different astrophysical environments are continuously exposed to energetic processing by photons, electrons, and ions that in some circumstances are particularly intense, for instance, in the magnetospheres of giant planets, within which many icy satellites are embedded. As a result, a variety of new chemical species are formed upon irradiation of icy dust mantles in the interstellar medium (ISM) and ice surfaces of the Solar System. On board the upcoming NASA James Webb Space Telescope (JWST) space mission, four instruments of unprecedented sensitivity (signal-to-noise ratio, S/N~$\sim100-300$) and spectral resolution (R~$\sim1500-3000$) over a wide wavelength range from the optical to the infrared ($0.6-28.3$~{\mc}, called NIRISS, NiRCam, NIRSpec, and MIRI) will systematically map ices across the sky by tracing many different evolutionary stages of star formation in space. A systematic set of vacuum ultraviolet-visible (VUV-vis) and near-infrared (NIR) laboratory spectra of the main interstellar ice species at appropriate temperatures will be particularly useful to aid radiative transfer modeling of protostars and disks observed by the JWST to better account for the amount of stellar radiation absorbed in the UV-vis and NIR by ices in the midplane of the disk \citep{McClure_2019}. Furthermore, a future ESA mission, the JUpiter ICy moons Explorer (JUICE), is set to make detailed observations of the giant gaseous planet Jupiter and three of its largest moons, Ganymede, Callisto, and Europa \citep{Banks2012}. On board the spacecraft, two instruments will cover the VUV-vis range: the UV imaging Spectrograph (UVS, $55-210$~nm with spectral resolution $\leq0.6$~nm), and the Moons and Jupiter Imaging Spectrometer (MAJIS, $400-5400$~nm with spectral resolution $3-7$~nm). This mission will be pivotal to our understanding of the Jovian system and can potentially unravel its history.

Because the two spectrometers on board the JUICE spacecraft will not cover the spectral region around 300~nm, a correct interpretation of the observational data can only be achieved by comparing these data to systematic VUV-UV-vis studies of laboratory analog materials in the full spectral range ($100-700$~nm). However, to date, laboratory studies of energetically processed ices are mostly carried out in the mid-infrared (MIR) spectral range by means of Fourier-transform infrared (FT-IR) spectroscopy \citep[e.g.,][]{Robert_Wu_etal2002, Palumbo_etal2008, MartinDomenech_etal2016, Hudson_2018}, and only a few are in the UV-vis \citep{Jones_etal2014b, Bouwman_etal2009}. Laboratory results on astrophysically relevant ices obtained in the VUV spectral range are equally sparse. To our knowledge, most  VUV spectroscopic studies of ices have explored pure molecules or simple mixtures \citep[e.g.,][]{Mason_etal2006, Sivaraman_etal2014, CruzDiaz_etal2014a, CruzDiaz_etal2014b, Wu_etal2012, Lo_etal2018, Lo_etal2019, Chou_etal2020}, and there have been only a few VUV photoabsorption studies of electron-irradiated ices in the spectral range $100-200$~nm \citep{Wu_etal2013a, Wu_etal2013b}.

Previous work has shown that molecules such as sulfur dioxide (SO$_{2}$) and ozone (O$_{3}$) can be potentially identified in Solar System objects in the VUV-UV-vis spectral range by comparing laboratory VUV-UV-vis data with available observational databases from previous missions such as Galileo, the Hubble Space Telescope, and Cassini \citep[e.g.,][]{Boduch_etal2016}. Here we present the first results of an ongoing experimental program aimed at systematically measuring VUV-UV-vis photoabsorption spectra in the range $115-700$~nm of ices irradiated with energetic (1\,keV) electrons under conditions relevant to the icy bodies of the Solar System and ice mantles of interstellar grains. Complementary FT-IR spectra of the same ices are also acquired to aid in the interpretation of the results. Our ultimate goal is to provide systematic VUV data that will help to identify spectral features such as absorption bands and/or spectral slopes as observed on icy objects in the Solar System by previous space-borne missions and the upcoming JUICE spacecraft. Our data will also help interpret JWST observations of star-forming regions in the ISM by providing valuable information on the VUV photoabsorption spectra of pure and mixed molecular ices as a function of temperature and exposure dose to 1\,keV electrons. Finally, VUV photoabsortion spectra of pure molecules will aid in the interpretation of photodesorption studies of the same species under space conditions.

This first study focuses on the VUV photoabsorption spectra of nitrogen oxides (N$_x$O$_y$) both deposited and formed upon 1\,keV electron irradiation of mixed homonuclear species, that is, the simplest astronomically relevant ice analogs. We present VUV-UV spectra ($115-340$~nm) of  molecular oxygen (O$_{2}$), nitrogen (N$_{2}$), and a 1:1 mixture over the temperature range $22-50$~K before and after 1\,keV electron bombardment. VUV absorption spectra of electron-irradiated ices of nitrogen oxide species, pure and combined in binary mixtures, are also presented for the first time. The choice of studying nonpolar homonuclear molecules was made because N$_{2}$ and O$_{2}$ are the most abundant gas-phase species in the Earth's atmosphere and are relevant to life. Moreover, they are observed in various environments in space \citep[e.g.,][]{Elsila_etal1997, Caselli_etal2002, Bieler_etal2015}, and their surface chemistry can potentially play an important role in the formation of nitrogen oxides as well as other more complex species \citep[e.g.,][]{Boduch_etal2012, Sicilia_etal2012, Vasconcelos_etal2017}.

Although molecular oxygen and nitrogen are abundantly present in the Solar System, their elusive presence in the ISM can be explained by the fact that both species are symmetric diatomic molecules with no allowed rotational or vibrational dipole transitions. Unlike most of the observed species in the ISM, molecular oxygen and nitrogen can therefore not be directly detected through either millimeter-wavelength observations of rotational emission lines or infrared spectroscopic detection of absorption or emission vibrational bands. Molecular nitrogen has been detected in diffuse interstellar clouds through its spectral lines, however, which are created by electronic transitions in the molecule at far-ultraviolet wavelengths toward the star HD 124314 by the Far Ultraviolet Spectroscopic Explorer (FUSE) mission \citep{Knauth_etal2004}. Upper limits for molecular O$_2$ in the ISM were found by a deep search toward low-mass protostars, NGC~1333–IRAS~4A (O$_2$/H$_2$~$<6\times10^{-9}$), by means of the Herschel Space Observatory  \citep{Yildiz_etal2013}.

In the ISM, gaseous O$_2$ and N$_2$ are thought to form primarily through the barrierless neutral-neutral reactions $\textrm{O}+\textrm{OH}\rightarrow{\textrm{O}}_2+\textrm{H}$ and $\textrm{N}+\textrm{NH}\rightarrow{\textrm{N}}_2+\textrm{H}$ (alternatively, $\textrm{N}+\textrm{OH}\rightarrow{\textrm{NO}}+\textrm{H}$ followed by $\textrm{N}+\textrm{NO}\rightarrow{\textrm{N}}_2+\textrm{O}$), respectively, in cold and warm  environments \citep{Pineau_etal1990, Hincelin_etal2011}. However, `nonenergetic' O$_2$ and N$_2$ surface formation reaction pathways on cold interstellar grains in dark clouds are also possible at low temperatures \citep[$10-20$ K;][]{Tielens_etal1982}. Laboratory and modeling work have shown that the chemistry involving O$_2$- and N$_2$-rich ices leads to the formation of O- and N-bearing inorganic and organic species throughout the star formation process \citep[e.g.,][]{Herbst_etal2009, van_Dishoeck_etal2013}. Although N$_2$ is mostly inert to nonenergetic atom addition reactions, the surface hydrogenation of O$_2$ ice forms solid water, the most abundant species detected on interstellar ice grains \citep[e.g.,][]{Ioppolo_etal2008, Ioppolo_etal2010, Miyauchi_etal2008, Matar_etal2008, Boogert_etal2015}. It is still debated, however, whether solid O$_2$ and N$_2$ observed in the Solar System have an interstellar origin.

Among the many important results  obtained by the Rosetta Orbiter mass Spectrometer for Ion and Neutral Analysis (ROSINA) was the measurement of  very high ratios of O$_2$/H$_2$O in the coma of comet 67P/Churyumov-Gerasimenko. This ratio fluctuates from 1 to $10\%$  and has a mean value of 3.8$\pm0.85\%$ \citep{Bieler_etal2015}. In addition, \cite{Rubin_etal2015} reanalyzed old measurements obtained for  1P/Halley by the Neutral Mass Spectrometer on board Giotto and found a value of 3.7$\pm1.7\%$ for the same ratio. A high abundance of O$_2$ therefore seems to be common in comets.

This result raised the question of the origin of molecular oxygen observed in cometary comae. Three scenarios have so far been considered. \cite{Mousis_etal2016} suggested that the high abundance of molecular oxygen is primordial and that a post-formation processing of cometary materials \citep[e.g., see][]{Strazzulla_Johnson1991} is not considered relevant. Along the same line of thought are the results obtained by \cite{Taquet_etal2016}, who developed models of interstellar gas-grain chemistry that favor a primordial origin for O$_2$ in comets, but for dark clouds, which are denser than the clouds that are usually considered as Solar System progenitors. The third scenario, by \cite{Dulieu_etal2017}, is different and suggests that molecular oxygen observed in the cometary comae is a daughter molecule being produced by the dismutation of hydrogen peroxide (H$_2$O$_2$).

Laboratory experiments show that irradiation of molecular oxygen-rich ice produces a large amount of solid ozone \citep[e.g., ][]{Cooper_etal2008, Jones_etal2014a}. However, only an upper limit for ozone detection was found in the coma of comet 67P/Churyumov-Gerasimenko \citep[$\leq2.5\times10^{-5}$ with respect to O$_2$; ][]{Taquet_etal2016}. This result can be explained either when we assume that the comet ice is largely unprocessed, preserving its pristine content of molecular oxygen, or when we consider that ozone can react, for example, by thermal processing with other surrounding molecules, and is then destroyed \citep{Strazzulla_etal2005, Loeffler_Hudson2016}.

N$_2$ has also been detected in the coma of comet 67P/Churyumov-Gerasimenko, but its abundance with respect to water is much lower, with a N$_2$-to-H$_2$O ratio of $\sim0.1\%$ \citep{Rubin_etal2015}. This value implies that nitrogen is depleted by a factor of about 25 with respect to the pre-solar level. \cite{Rubin_etal2015} suggested that this is due to inefficient trapping of N$_2$  in amorphous water ice and implies a formation temperature of $24-30$~K (i.e., below 24~K, the efficiency of nitrogen trapping would increase, and above 30~K it is no longer trapped), as evidenced by laboratory experiments using mass spectrometry after thermal desorption of appropriate mixtures \citep{BarNun_etal2007}. The same authors \citep{Rubin_etal2015} did not exclude that the measured ratio  may reflect in whole or in part the cometary post-formation evolution, however. In this scenario, comets could be formed in the trans-Neptunian region at temperatures lower than 24~K, and they would have trapped  much more nitrogen \citep[ e.g., as is observed to be the case for Pluto and Triton;][]{Owen_etal1993, Schmitt_etal2017} that decreased with time due to  the radiogenic decay of nuclides \citep[or to post-accretion energetic processing, e.g., see][]{Strazzulla_Johnson1991} and thermal cycles during their transit near the Sun.

Mixed with methane, molecular nitrogen is also present in aerosols in the upper atmosphere of the Saturn moon Titan \citep{Samuelson_etal1997, Niemann_etal2005}. N$_2$ has been detected at the surface of Triton, Pluto, and several Kuiper belt objects (KBOs) such as Eris and Makemake \citep{Cruikshank_etal1993, Owen_etal1993, Licandro_etal2006, Cruikshank_etal2015, Grundy_etal2016}. N$_2$ is further observed along with other nitrogen-bearing molecules such as ammonia and HCN in the gas phase in the plumes of Enceladus with an abundance of about $1\%$ with respect to water \citep{Waite_etal2009}. Plumes have also been detected on Europa through observations of water-molecule ejection, and it is assumed they might have a composition similar to that of Enceladus \citep{Teolis_etal2017}. Laboratory results from different independent groups corroborate the hypothesis that if molecular nitrogen is a component of the ice surfaces of the Solar System satellites and comets, then the products of its radiolysis in an oxygen-rich environment, that is, nitrogen oxides, should be searched for in space \citep{Jamieson_etal2005, Sicilia_etal2012, de_Barros_etal2015, Vasconcelos_etal2017, Hudson_2018}.

The unambiguous detection, or nondetection, of solid N$_x$O$_y$ species in space could give insights into the chemical processes occurring at the surfaces of Pluto, Triton, and 67P/Churyumov-Gerasimenko. If nitrogen oxides are observed on Pluto, it might be indirect proof of the presence of molecular oxygen. The lack of abundant oxygen and of nitrogen oxides currently favors the permanence of a high nitrogen abundance. The putative presence of N$_x$O$_y$ molecules in comets might also help to understand the fate of nitrogen and contribute to the debate of its original abundance. To date, a comprehensive selection of VUV laboratory spectra of nitrogen oxides, deposited and formed upon energetic processing of their parent species, in support of their future observations in the Solar System in the VUV-UV-vis spectral range is still lacking.

In this work, we present the first systematic set of VUV data of 1\,keV electron-irradiated oxygen- and nitrogen-bearing ices starting from pure O$_2$ and N$_2$ ices, followed by the irradiation of the binary O$_2$:N$_2$~=~1:1 mixture. We furthermore investigate the 1\,keV electron exposure of pure and mixed nitrogen oxides to study their formation and destruction pathways and their specific chemical stability upon energetic processing to inform future observations. In the next section we present the experimental apparatus in detail. Results and discussion first focus on VUV spectra of pure ices and then shift to ice mixtures before and after 1\,keV electron irradiation and thermal heating. Infrared spectra of selected ices are shown to retrieve complementary information. Astrophysical implications highlight the importance of a deep search for nitrogen oxide ices in space.

\section{Experimental setup}

\begin{figure*}
\centering
\hspace{-10.5mm}
\includegraphics[width=0.9\textwidth]{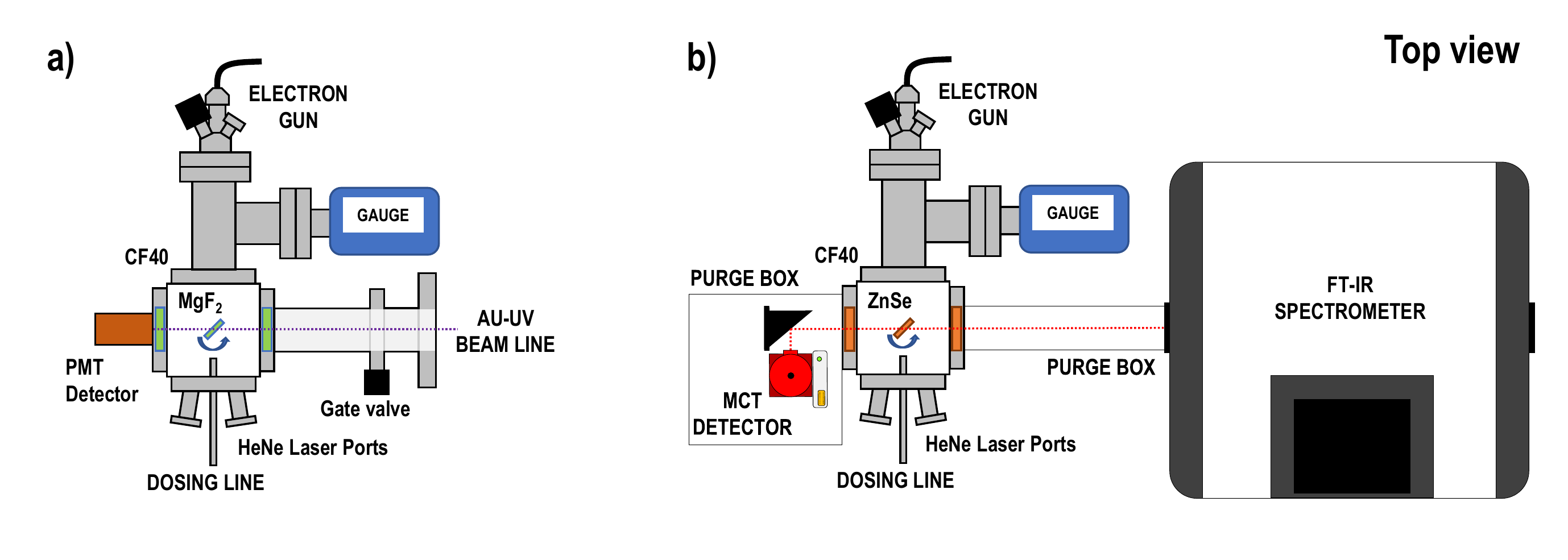}
\vspace{1mm}
\caption{Schematic top view of the Portable Astrochemistry Chamber (PAC) used to measure VUV and MIR spectra of astrochemical ices at a) ASTRID2 synchrotron light source at Aarhus University in Denmark, and b) the Molecular Astrophysics Laboratory at The Open University, UK, respectively.}
\label{Schematic}
\end{figure*}

Experiments were performed at two different sites using a custom-made Portable Astrochemistry Chamber (PAC), which has a base pressure of 10$^{-9}$~mbar. VUV photoabsorption spectra were measured on the AU-UV beam line at the ASTRID2 synchrotron light source at Aarhus University in Denmark \citep[Figure~\ref{Schematic}a; see][]{Eden_etal2006, Palmer_etal2015}. The beam line can produce monochromatized light over the wavelength range 115 to 700~nm using two gratings. The high-energy grating used for the measurements presented here has a typical flux of 10$^{10}$~photons/s/100~mA and a photon resolution of 0.08~nm. FT-IR spectra of ices produced using the same PAC were measured at the Molecular Astrophysics Laboratory of The Open University (OU), UK \citep[Figure~\ref{Schematic}b; see][]{Dawes_etal2016}. The advantage of using the same chamber for both types of measurements is that experimental conditions and preparation of ices can be accurately reproduced. Moreover, combining VUV and FT-IR spectroscopy of homonuclear species that are generally infrared transparent allows for their detection in the VUV, while their electron irradiation products can be monitored in the VUV and MIR.

The PAC consists of a conflat (CF40) flange spherical cube (Kimball Physics) connected to a 300~l/s turbo molecular pump (Leybold), a closed cycle helium cryostat (Sumitomo) with a base temperature 20~K, and a 1\,keV electron gun (Kimball Physics). For measurements at ASTRID2, the chamber is mounted on the exit port of the AU-UV beam line and is enclosed with magnesium fluoride (MgF$_{2}$, Crystran) windows to allow transmission of the VUV synchrotron light through the substrate and chamber; the transmitted light intensity is measured using a photomultiplier tube (PMT, ET enterprises 9406B). VUV photoabsorption spectra are acquired in transmission mode and at 1~nm intervals in the spectral range $115-340$~nm. During VUV photoabsorption measurements, the space between the MgF$_{2}$ exit window and the PMT is evacuated to remove atmospheric gas signatures at low wavelengths ($<200$~nm), while at longer wavelengths, the atmospheric O$_2$ prevents the transmission of second-order light. Ices are prepared on a 25~mm diameter MgF$_{2}$ window (Crystran); this substrate is mounted in a holder made of oxygen-free high-conductivity copper (OFHC, Goodfellow), which has been machined to give good thermal contact between the substrate and the cryostat. The temperature of the substrate is measured with a DT-670 silicon diode (Lakeshore) and can be controlled in the range $20-300$~K by means of a Kapton tape heater (Omega) connected to the OFHC block and regulated with a temperature controller system (Oxford Instruments). For FT-IR measurements carried out at the OU, the MgF$_{2}$ windows are replaced with zinc selenide (ZnSe, Crystran), which are $\sim90\%$ transparent in the MIR. The FT-IR spectrometer used is a Nicolet Nexus 670, with the light transmitted through the system detected using a mercury cadmium telluride (MCT) detector. FT-IR spectra are acquired in transmission mode, with absorbance spectra collected at 1~cm$^{-1}$ resolution with an average of 128 scans in the range $4000-600$~cm$^{-1}$. Purge boxes visible in Figure~\ref{Schematic}b are connected to a dry compressed air system and are used to remove atmospheric gas-phase signatures along the MIR beam path outside the high-vacuum chamber.

During a standard experiment, the substrate is first cooled down to 20~K, flash-heated to 200~K, and then cooled back to 20~K to remove possible contaminants from the surface. When the substrate is at its selected temperature for the experiment, a background spectrum is acquired and used as a reference for sample spectra taken afterward. Pure or mixed ($\geq99.999\%$ N$_{2}$, $\geq99.998\%$ O$_{2}$, $98.5\%$ NO, $99\%$ N$_{2}$O, and $99\%$ NO$_{2}$; CANgas by Sigma-Aldrich) gases are prepared using a dedicated gas line, with ratios determined through partial pressures measured with a mass-independent baratron (MKS) prior to deposition. Gases are then admitted from the mixing gas line to the main chamber through an all-metal needle valve. During deposition, the pressure is controlled and regulated between $1\times10^{-8}$ and $1\times10^{-7}$~mbar by means of a mass-dependent ion gauge (Leybold IONIVAC ITR 90), which gives deposition rates of between 0.02 and 0.2~nm~s$^{-1}$. Deposition is carried out at normal incidence with respect to the substrate. A rotary stage mounted in between the main chamber and the cryostat head allows for the rotation of the substrate so that it can be positioned with its surface normal to either 1) the incident light, when a spectrum is acquired; 2) the inlet tube of the gas deposition line during deposition of gases; or 3) the electron beam during irradiation (see Figure~\ref{Schematic}). It should be noted that nitrogen dioxide (NO$_{2}$) is always deposited in both its monomeric and dimeric (N$_{2}$O$_{4}$) forms. When we refer to NO$_{2}$ ice, we therefore imply an NO$_{2}$:N$_{2}$O$_{4}$ ice mixture.

The thickness of the deposited ice samples is determined using a laser-interference technique described in detail elsewhere \cite[e.g.,][]{Born_Wolf1970, Goodman1978, Baratta_Palumbo1998}. The technique is based on monitoring the intensity variations of a Ne-He laser beam reflected off the sample surface with an angle of incidence of 20$^{\circ}$ from the normal to the surface during ice-film growth. The intensity of reflected laser light is measured using a silicon photodiode (ThorLabs). As the ice layer is growing, a sinusoidal variation in intensity is detected due to the interference of laser light reflected from the ice surface-vacuum and the sample-substrate interface. The thickness of the ice layer is then calculated using Equation \ref{thickness},

\begin{equation}
                l~=~{\frac{\lambda_{0}}{2n_{1}\cos\theta_{1}}}\times{N}
        \label{thickness}
,\end{equation}

\noindent
where $\lambda_{0}$ is the wavelength of the Ne-He laser beam  in vacuum (632.8~nm), $\theta_{1}$ is the angle of the laser within the ice, $n_{1}$ is the refractive index of the ice film, and $N$ is the number of constructive pattern repetitions during the deposition time. The refractive index $n_{1}$ at 632.8~nm is estimated from the ratio of the maxima and minima of the laser-interference pattern \citep{Born_Wolf1970, Berland_etal1994, Westley_etal1998}. The refractive indices of pure ices and their mixtures are estimated to be 1.33 for both pure molecular oxygen and nitrogen, and 1.34 for the O$_{2}$:N$_{2}$~=~1:1 mixture (see Table~\ref{experiments}). The value of refractive index of the MgF$_{2}$ substrate (1.38) is taken from \cite{Dodge1984}. The ices presented in this work are in the range $0.11-1.35$~\mc\ thick.

After deposition, measurements are carried out to provide an absorbance spectrum of the unirradiated sample. Samples are then irradiated using a 1\,keV electron gun. The electron-beam current of the 1\,keV gun was measured at the center of the chamber with a Faraday cup placed instead of the substrate holder, and the flux of electrons is derived to be $2\times10^{13}$~e$^{-}$cm$^{-2}$s$^{-1}$ through the following equation:

\begin{equation}
                \Phi_{\textrm{1\,keV e-}}~=~{\frac{I}{e \times A}}
        \label{flux}
,\end{equation}

\noindent
where $\Phi_{\textrm{1\,keV e-}}$ is the 1\,keV electron flux [e$^{-}$cm$^{-2}$s$^{-1}$], $I$ is the electron beam current ($\sim11~\mu{\textrm{A}}$) measured by the Faraday cup, $e$ is the elementary charge ($\sim1.60\times10^{-19}$ C), and $A$ is the surface area of the substrate exposed to the electron beam ($\sim3.14$~cm$^2$).

\begin{figure}
\centering
\includegraphics[width=0.5\textwidth]{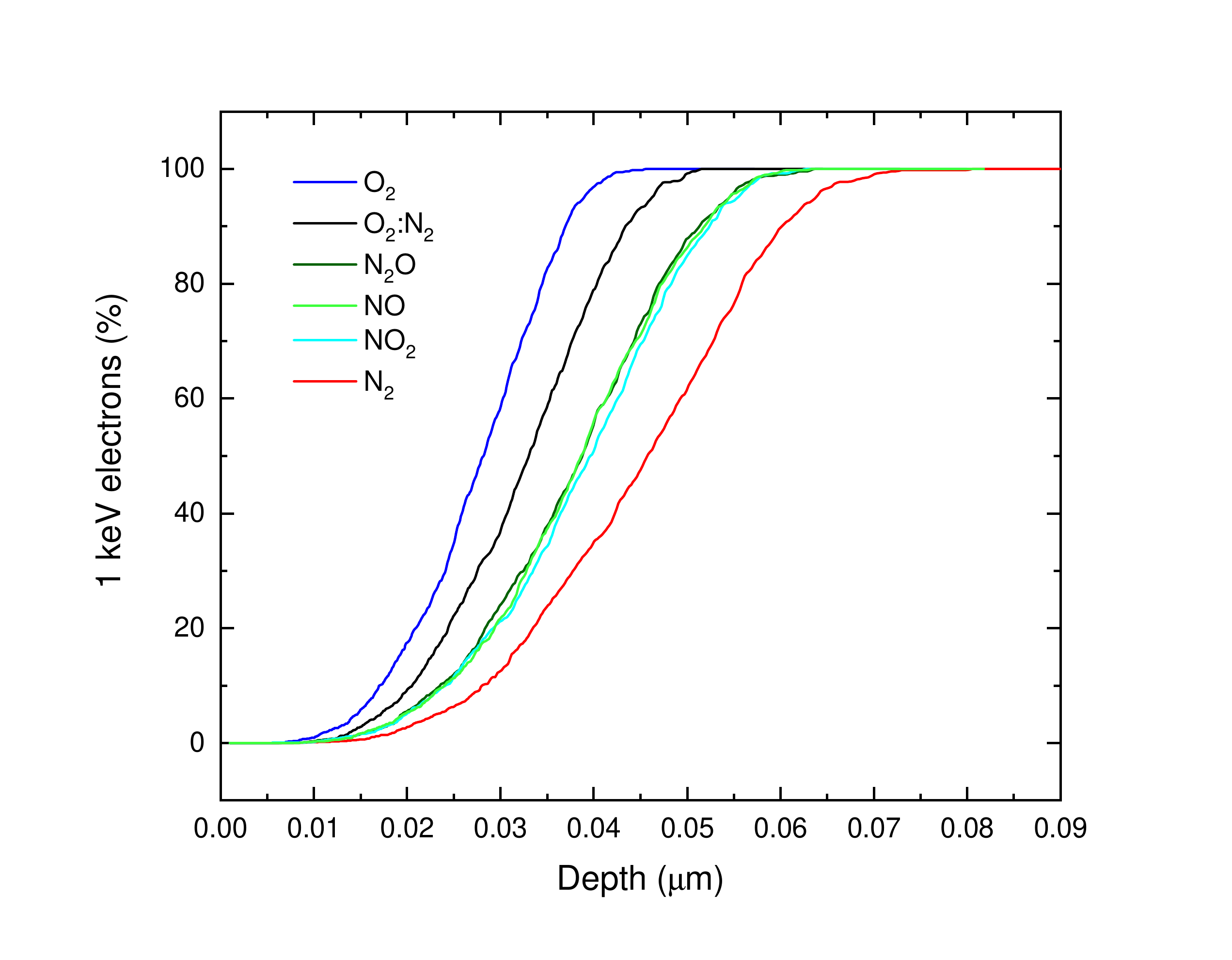}
\vspace{-5mm}
\caption{Results from CASINO simulations of 1\,keV electrons trajectories in N- and O-rich ices. For each simulation 1000 trajectories were calculated. All electrons impinging normal to the O$_2$ ice surface are stopped at 0.045~\mc\ (blue line), while only about 50, 70, and $90\%$ of electrons are stopped in solid N$_2$ (red line), nitrogen oxides (light and dark green and light blue lines) and the O$_2$:N$_2$~=~1:1 ice mixture (black line) at this depth.}
\label{PD}
\end{figure}

The penetration depth of 1\,keV electrons is estimated using the CASINO software \citep[Monte Carlo simulation of electron trajectory in solids;][]{Drouin_etal2007} to be 0.045~\mc\ in the case of pure O$_{2}$, 0.07~\mc\ for pure N$_{2}$, and 0.06~\mc\ for pure nitrogen monoxide (NO), nitrous oxide (N$_{2}$O), and NO$_{2}$ ices (see Figure~\ref{PD}). The penetration depth for the O$_{2}$:N$_{2}$~=~1:1 mixture is 0.05~\mc, assuming an intermediate density between those of O$_{2}$ and N$_{2}$. For all mixtures involving NO, N$_{2}$O, and NO$_{2}$ molecules, a penetration depth of 0.07~\mc\ is assumed. We further verified experimentally an upper limit for the penetration depth of 1\,keV electrons in O$_2$ ice by performing a control experiment in which we first deposited 0.7~\mc\ N$_2$ ice at 22~K and then 0.14~\mc\ O$_2$ ice on top of it at the same temperature. Further electron exposure of this layered ice produced O$_3$ only, which is an O$_2$ irradiation product. Except for N$_2$, no traces of other N-bearing species were observed. This confirms that the values of the penetration depth of 1\,keV electrons in O$_2$ ice as calculated by CASINO are reliable. Thus, since all the ice thicknesses of pure and mixed O$_{2}$ and N$_{2}$ presented in this study are between 0.1 and 1.3~\mc\, all impinging electrons are implanted in the ice upon irradiation. For an overview of the experimental parameters, see Table~\ref{experiments}.

When electron irradiation is completed and a spectrum is measured, the sample temperature is then increased to 30, 35, 50, 80, and 100~K, recording a new absorbance spectrum at each temperature. Experiments measuring the VUV absorption cross section of pure NO, N$_{2}$O, and NO$_{2}$ ices deposited at 22~K are also performed to aid the spectral identification of nitrogen oxides in the VUV range upon 1\,keV electron exposure of N- and O-rich ices.

\begin{table*}
\centering
        \caption{List of main experimental parameters, i.e., density of material, refractive index $n_{1}$ at 632.8~nm, ice thickness $d$, maximum fluence, and dose. For each experiment, we provide a list of all new species observed in the VUV range.}
        \label{experiments}
        \begin{tabular}{lllllll} 
                \hline
                Sample & Density &  $n_{1}$ & $d$ & Fluence & Dose & New species observed \\
        & g/cm$^{3}$ & & \mc & 10$^{16}$ e$^-$/cm$^{2}$ & eV/16u & in the VUV range\\
                \hline
                O$_2$                  & 1.54$^{b}$ & 1.33 & 0.25 & 7.2 & 277.6 & O$_3$\\
                N$_2$                  & 0.94$^{a}$ & 1.33 & 1.35 & 43 & 1746.1 & N$_3$\\
                O$_2$:N$_2$~=~1:1        & 1.24$^{c}$ & 1.34 & 0.24 & 2.4 & 104.3   & O$_3$, N$_2$O$^e$, NO$_2$\\
                NO                     & 1$^{d}$    & 1.41 & 0.12 & 1.8 & 80.2  & O$_2^e$, NO$_2$\\
                N$_2$O                 & 1.16$^{b}$ & 1.39 & 0.12 & 0.7 & 27.6  & NO$_2$\\
                NO$_2$                 & 1.17$^{b}$ & 1.44 & 0.11 & 2.4 & 91.4  & O$_2^e$\\
                NO:NO$_2$~=~1:1          & 1.1$^{c}$  & 1.42 & 0.11 & 3.9 & 157.9 & ---\\
        NO:N$_2$O~=~2:1          & 1.1$^{c}$  & 1.39 & 0.12 & 3.6 & 145.7 & O$_2^e$, N$_2^e$, NO$_2$ \\
                NO$_2$:N$_2$O~=~1:1      & 1.1$^{c}$  & 1.39 & 0.12 & 3.6 & 145.7  & ---\\
                \hline
        \end{tabular}
\begin{flushleft}
\footnotetext{}{$^{(a)}$\cite{Satorre_etal2008}; $^{(b)}$\cite{Fulvio_etal2009}; $^{(c)}$estimated average value from mixture components; $^{(d)}$no value available in the literature, hence assumed to be 1; $^{(e)}$tentative detection.}
\end{flushleft}
\end{table*}

\section{Results and discussion}

Very much in line with other laboratory work on radiation-driven oxidation of nitrogen-bearing ices \citep[e.g., ][]{Hudson_2018}, our laboratory results confirm that upon irradiation of oxygen- and nitrogen-bearing ices, new species are formed, such as O$_{3}$, azide (N$_{3}$), NO, N$_{2}$O, and NO$_{2}$. We discuss the results for irradiation of each ice in turn. Table~\ref{experiments} lists all new species observed in the VUV range upon 1\,keV electron irradiation of N- and O-rich ices.

\subsection{Pure O$_{2}$ ice}
\begin{figure}
\centering
\includegraphics[width=0.5\textwidth]{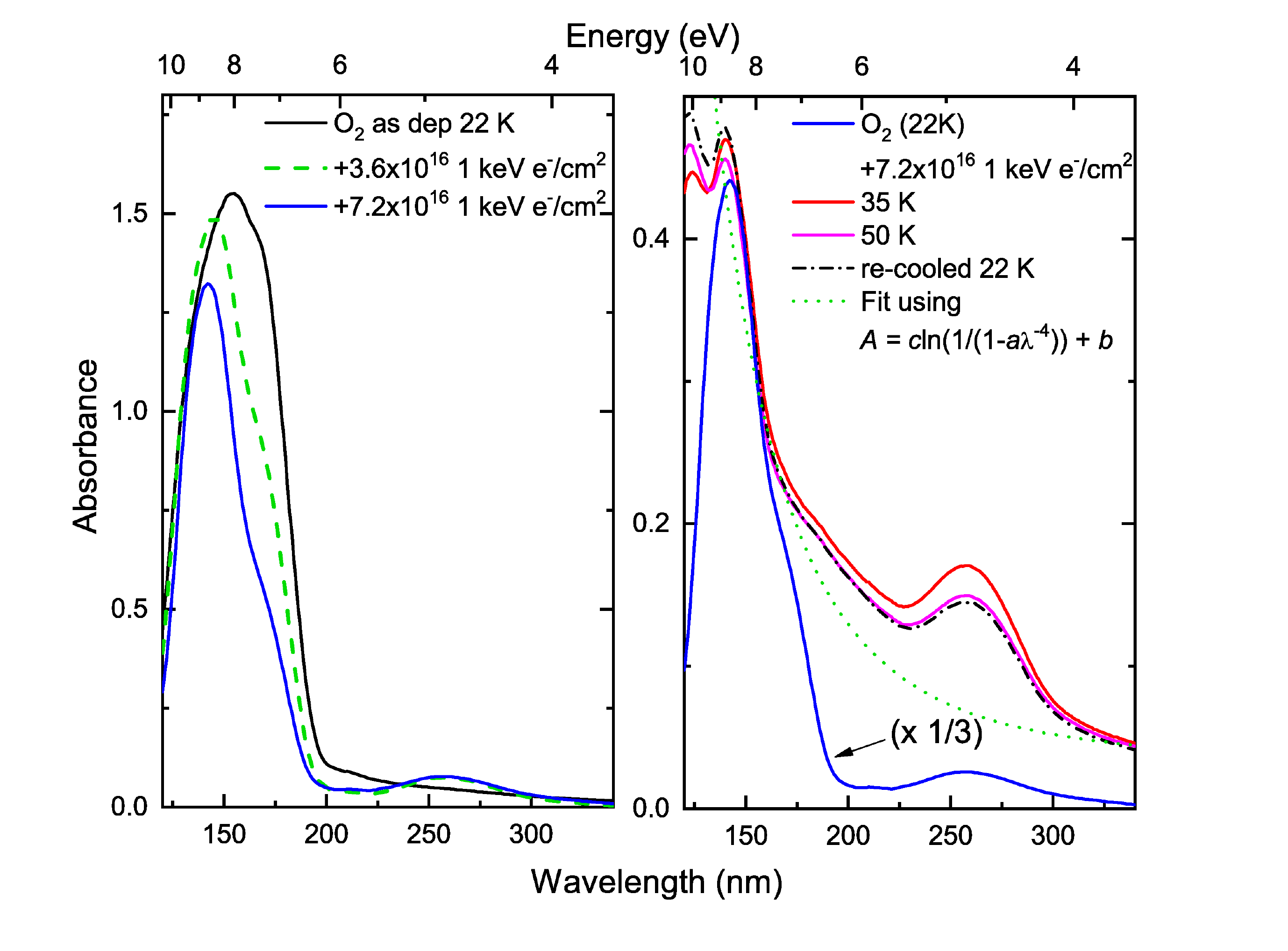}
\vspace{-5mm}
\caption{Left panel: VUV absorbance spectrum ($120-340$~nm) of frozen oxygen as deposited at 22~K (solid black curve) and after 1\,keV electron irradiation at two different fluences (dashed green and solid blue lines). Right panel: VUV absorbance spectrum ($120-340$~nm) of electron-irradiated frozen oxygen (22~K; solid blue line) is shown along with the spectra obtained after heating to 35~K (solid red line), 50~K (solid magenta line), and then cooling down to 22~K (dot-dashed black line). The dotted green line is a fit of the 35~K spectrum by means of Equation~\ref{slope} to highlight the Rayleigh-scattering tail in the spectrum.}
\label{O2}
\end{figure}

The left panel of Figure~\ref{O2} shows the VUV photoabsorption spectrum of 0.25~\mc\ thick pure O$_{2}$ deposited at 22~K (black curve). \cite{Mason_etal2006} showed a similar absorption spectrum of solid O$_{2}$ and identified the 140~nm feature to be the Schumann-Runge band (S-R, \textrm{$\widetilde{B}^{3}\Sigma^{-}_{u}\leftarrow\widetilde{X}^{3}\Sigma^{-}_{g}$} transition from the ground state) assigned from gas-phase data. In their work the second and less intense solid O$_{2}$ band at 180~nm was suggested to be due to the oxygen dimer (O$_{2}$)$_{2}$. In the O$_{2}$ spectrum of Figure~\ref{O2} these bands appear to be slightly redshifted, most likely because the spectrum is saturated by the excessive thickness of the ice in combination with a high absorption cross section of O$_{2}$ ice in the VUV. Thick ices were deposited to ensure that the penetration depth of the impinging electrons would not exceed the thickness of the ice.

In the left panel of Figure~\ref{O2} we show along with the spectrum of pure O$_{2}$  two spectra of the same ice obtained after 1\,keV electron irradiation at two different fluences (3.6 and 7.2$\times$10$^{16}$\,e$^-$/cm$^{2}$). When a density of 1.54~g/cm$^{3}$ \citep{Fulvio_etal2009} is considered for the O$_{2}$ ice, the average stopping power of an impinging electron, that is, the energy released by 1\,keV electrons into O$_{2}$ ice, is 2.22~$\times$10$^{4}$~eV/\mc, or in other units, 3.9$\times~10^{-15}$~eV~cm$^{2}$/16u, where u is the unified atomic mass unit defined as 1/12 of the mass of an isolated atom of $^{12}$C. Because the sample was irradiated for a maximum of 7.2$\times$10$^{16}$\,e$^-$/cm$^{2}$, the total energetic dose is 277.6~eV/16u.

As shown in Figure~\ref{O2}, the 1\,keV electron irradiation of the pure O$_{2}$ layer causes several modifications in the ice, highlighted by changes in the VUV spectral profile at 22~K. First, the O$_{2}$ absorption bands decrease in intensity because the desorption of the ice induced by the impinging electrons is efficient. Another mechanism that leads to O$_{2}$ consumption is the formation of solid ozone \citep{Bennett_Kaiser2005}. Second, the O$_{2}$ band peaks shift to higher energies. The peak shifts are a sign that the spectra of the irradiated ice are no longer saturated because the ice is now thinner than immediately after deposition. The O$_{2}$ band profiles and peak positions of the irradiated ice correspond to those published in \cite{Mason_etal2006} and \cite{Sivaraman_etal2014}. Third, a new absorption band appears at 257~nm. This is the Hartley band of solid ozone (O$_{3}$) formed upon the reaction $\textrm{O}_{2}(\widetilde{X}^{3}\Sigma^{-}_{g})+\textrm{O}(^{3}P)\rightarrow{\textrm{O}_{3}(\widetilde{X}^{1}A_{1})}$, where atomic oxygen is liberated in the ice upon the interaction of 1\,keV electrons with O$_{2}$ molecules \citep{Bennett_Kaiser2005, Sivaraman_etal2014}.

The right panel of Figure~\ref{O2} shows the spectrum of the final irradiation of the O$_{2}$ ice at 22~K scaled to one-third of its original absorbance for clarity (solid blue line) and compared to the spectra of the same ice annealed to 35~K (solid red line), 50~K (solid magenta line), and then cooled again down to 22~K (dash-dotted black line). At 35~K, most of the O$_{2}$ has left the ice through thermal desorption, leaving a layer of more or less pure O$_{3}$ ice, with peaks at 140 and 257~nm. \cite{Sivaraman_etal2014} assigned these peaks to ozone ice. However, the 140~nm peak could also be assigned to O$_2$ molecules trapped in an O$_3$ ice matrix (see Fig.~\ref{O2}). There is also a clear underlying slope in the VUV photoabsorption spectra of irradiated and annealed ices most likely due to Rayleigh scattering off the rough surface of the ice as a result of irradiation, e.g., sputtering and/or desorption of top material. Following \cite{Dawes_etal2018}, the scattering tails in the absorbance spectra of the irradiated and annealed O$_2$ ice were fit with a function of the form

\begin{equation}
                A~=~c\textrm{ln}\binom{1}{\overline{1-a\lambda^{-4}}} + b
        \label{slope}
,\end{equation}

\noindent
where the absorbance ($A$) Equation, which is first derived using the rearranged form of the Beer–Lambert Law, is then modified by introducing the term $a\lambda^{-4}$ in the denominator corresponding to loss in the transmitted intensity due to scattering. Here $a$ is proportional to $r^6$, where $r$ is the scattering particle size and $c$ is proportional to the number density of the scatterers in the beam path. The dotted green line in the right panel of Figure~\ref{O2} is the fit of the VUV photoabsorption spectrum at 35~K using Equation~\ref{slope}. The good agreement of the fit with the spectrum at 35~K in the spectral regions around 180 and 320~nm suggests that at least part of the observed slope is due to Rayleigh scattering. On the other hand, the large discrepancies between the fit and the VUV spectrum, especially around 257~nm, indicate that there are real absorption bands due to O$_3$ (e.g., at 257~nm) and possibly also to O$_2$ at 140~nm.

Furthermore, it appears that the ozone band profile changed at 35~K when it is compared to its profile after electron irradiation at 22~K. A possible reason for band profile changes is that the VUV spectrum of pure ozone presents a slightly different absorption cross section than when O$_3$ is diluted in an O$_2$ ice matrix. \cite{Bennett_Kaiser2005} showed that the ozone-oxygen complex [O$_3$...O] is formed upon exposure of O$_2$ ice to 5~keV electron irradiation at low temperatures ($11-30$~K). They verified experimentally that in the warm-up phase, the oxygen atoms from the [O$_3$...O] complex react with a neighboring oxygen molecule to form an O$_3$ dimer. Similarly, during warm-up, oxygen atoms trapped in the molecular oxygen matrix react with oxygen molecules to form the ozone monomer. Thus, further formation of ozone can also contribute to the observed changes in the profile of the 257~nm band at 35~K. At 50~K, the 257~nm ozone band changes again, which is a sign of crystallization, as pointed out by \cite{Sivaraman_etal2014}. It is worth noting that there are no major differences between the spectra at 50~K and the one acquired after cooling of the ice again  at 22~K, which indicates the irreversibility of the phase transition.

\begin{figure}
\centering
\includegraphics[width=0.5\textwidth]{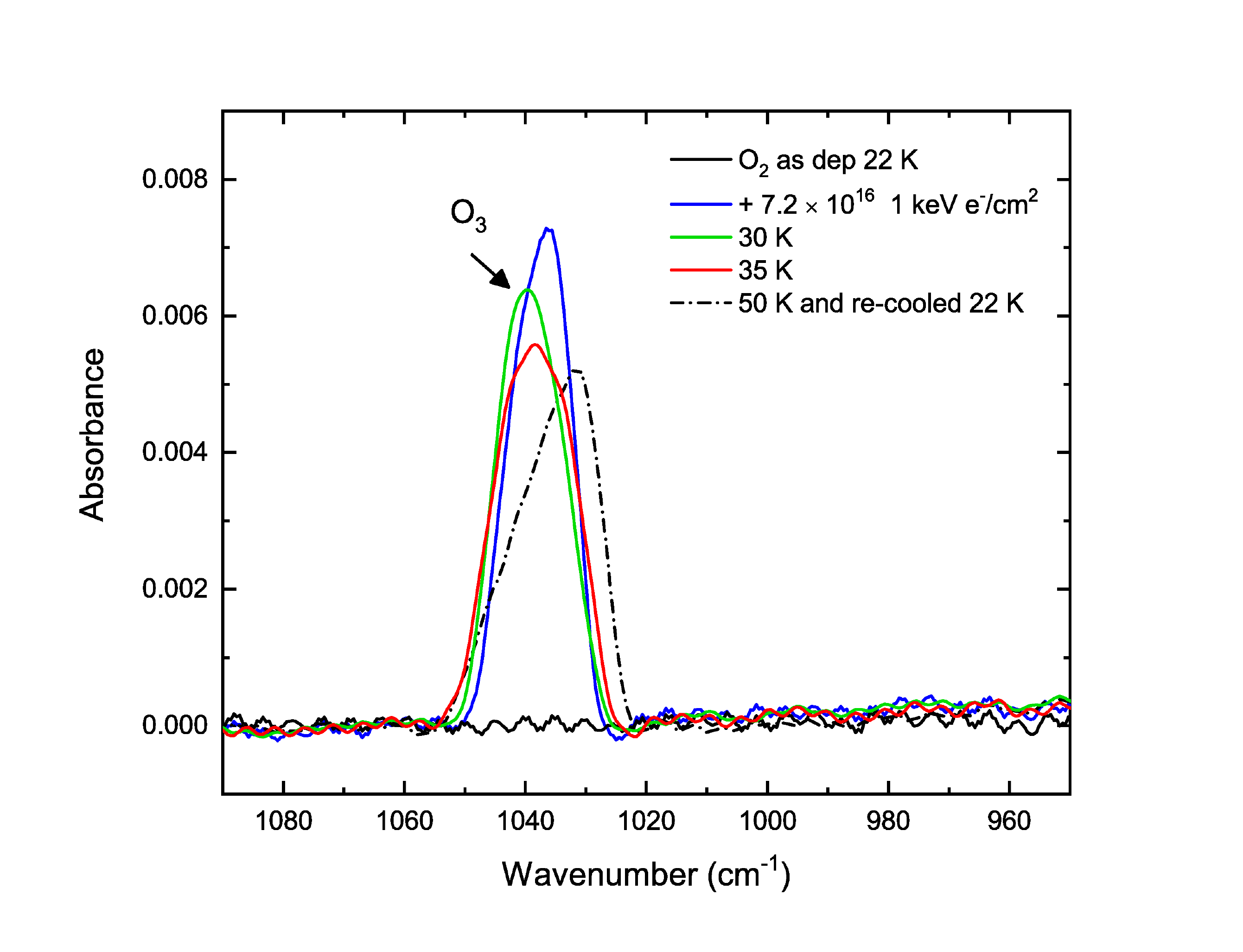}
\caption{FT-IR spectrum in absorbance of an O$_2$ ice as deposited at 22~K (solid black line), after 1\,keV electron bombardment (solid blue line), and annealed to 30~K (solid green line), 35~K (solid red line), and 50~K, then cooled to 22~K (dot-dashed black line) in the spectral region around the antisymmetric stretching mode of solid O$_3$. }
\label{O2_IR}
\end{figure}

Figure~\ref{O2_IR} shows the infrared spectra of O$_{2}$ at 22~K before and after exposure to 1\,keV electrons for the same total dose as we used in the experiments carried out at ASTRID2 and discussed above. After irradiation, the previously featureless MIR spectrum of pure molecular oxygen now shows an absorption band due to ozone formed in the ice at 22~K \citep{Bennett_Kaiser2005}. The peak position and profile change when the irradiated ice is annealed to 30, 35, and 50~K and then cooled again to 22~K. As discussed before, spectral changes arise because heating allows molecular oxygen to desorb, leaving a layer of pure ozone. The blueshift of the ozone peak position seen in the MIR spectrum at 30~K is likely due to the formation of a pure layer of O$_3$ ice. However, at 50~K, the spectrum of ozone is now redshifted. Changes in the peak position and band profile at 50~K can be due to a restructuring of the ice layer. \cite{Sivaraman_etal2007} showed that a phase transition from $\alpha$ (amorphous) to $\beta$ (crystalline) ozone occurs when O$_3$ ice is heated to 47~K. At 50~K, O$_3$ is therefore crystalline, while its desorption occurs at about 60~K \citep{Jones_etal2014a}.

\subsection{Pure N$_{2}$ ice}
\begin{figure}
\centering
\hspace{-12.5mm}
\includegraphics[width=0.5\textwidth]{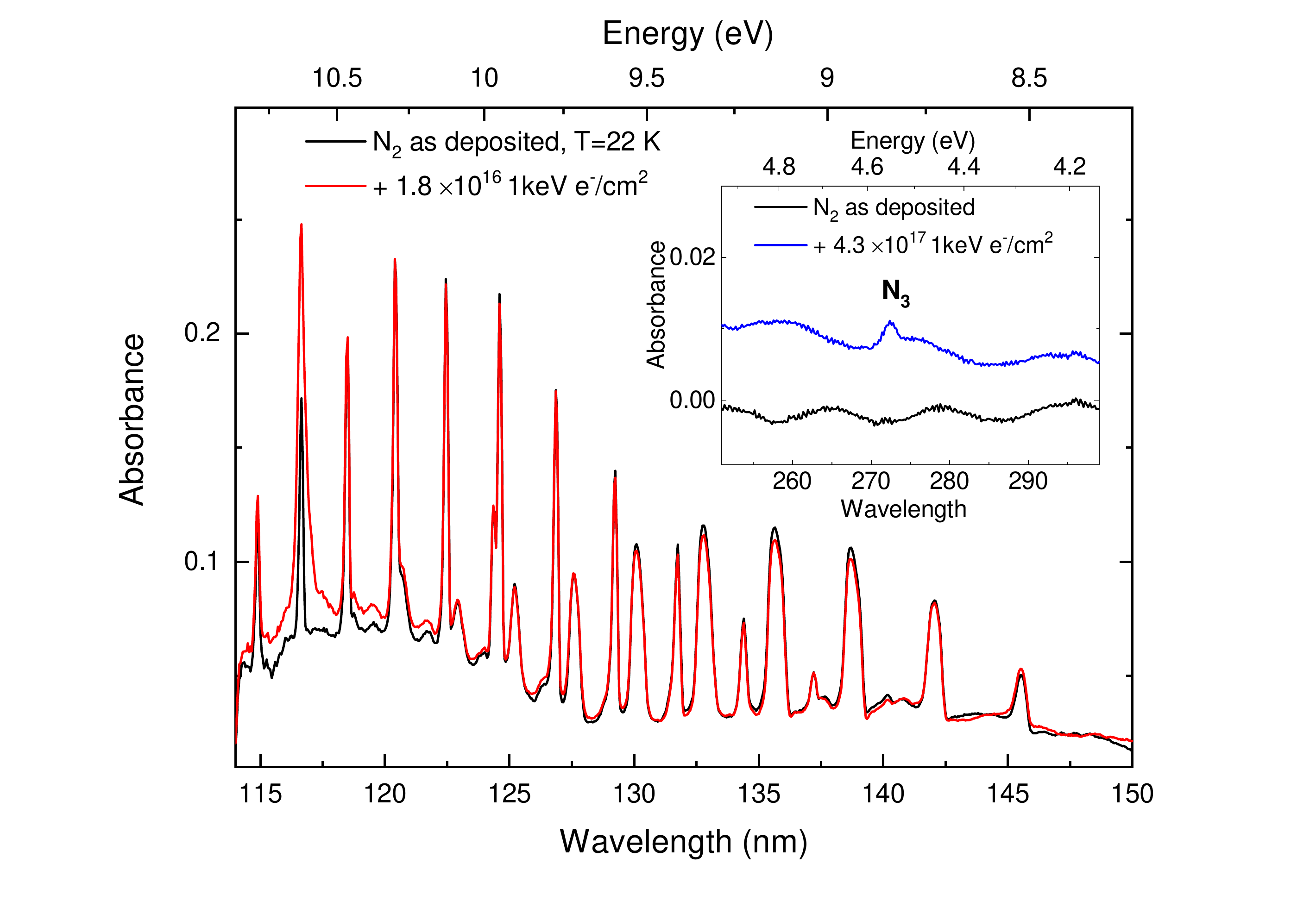}
\caption{VUV absorbance spectrum ($115-150$~nm) of frozen nitrogen as deposited at 22~K (solid black line) and after 1\,keV electron irradiation at given fluences (solid red and blue lines). Formation of N$_3$ is highlighted in the inset of the figure (solid blue line).}
\label{N2irr_inletN3}
\end{figure}

The absorbance spectrum ($115-150$~nm) of a $\sim$1~\mc\ thick N$_2$ frozen layer deposited at 22~K is shown in Figure~\ref{N2irr_inletN3} along with the spectrum obtained after 1\,keV electron irradiation with a fluence of 1.8$\times$10$^{16}$\,e$^-$/cm$^{2}$. To highlight the sharp features of N$_2$ in the VUV, the step size used here is 0.05 nm for the large panel of Figure~\ref{N2irr_inletN3}, while the inset is at a step of 0.2~nm. The spectrum of solid molecular nitrogen shows a series of intense absorption features in the range $115-150$~nm associated with the Lyman–Birge–Hopfield (LBH, \textrm{$a^{1}\Pi_{g}\leftarrow\widetilde{X}^{1}\Sigma^{+}_{g}$}) and Tanaka absorption (TA, \textrm{$w^{1}\Delta_{u}\leftarrow\widetilde{X}^{1}\Sigma^{+}_{g}$}) systems \citep{Mason_etal2006, Wu_etal2012}. It should be noted that to obtain the spectra shown in Figure~\ref{N2irr_inletN3} we deposited a much thicker ice layer than for the case of molecular oxygen because photoabsorption cross section for solid N$_2$ is much smaller. Moreover, with the exception of one single line at about 117\,nm, which increases in intensity after exposure to 1\,keV electrons, there are no major changes in the spectrum of solid N$_2$ in the range $115-150$~nm upon irradiation. Unlike for the case of pure O$_2$, 1\,keV electron exposure of a pure N$_2$ ice therefore leads to a negligible desorption of the ice. However, the origin of the band increase at 117~nm is unclear at this stage.

Formation of the N$_3$ (azide) radical is confirmed by the appearance of the peak at $\sim$272.4~nm (transition \textrm{$\widetilde{A}^{2}\Sigma^{+}_{u}\leftarrow\widetilde{X}^{2}\Pi_{g}$}) \citep[e.g.,][]{Douglas_Jones1965, Wu_etal2013b}, as highlighted in the inset of Figure~\ref{N2irr_inletN3}, where a spectrum of deposited N$_2$ ice and another spectrum after irradiation of 4.3$\times$10$^{17}$\,e$^-$/cm$^{2}$ are shown. Because the average stopping power of a 1\,keV electron impinging N$_2$ ice with a density of 0.94~g/cm$^{3}$ \citep{Satorre_etal2008} is estimated to be 4.05$\times 10^{-15}$~eV~cm$^{2}$/16u (1.42~$\times~10^{4}$~eV/\mc), a total fluence of 4.3~$\times$10$^{17}$~e$^-$/cm$^{2}$ translates into a dose of 1746.1~eV/16u. The laboratory formation of the N$_3$ radical initiated by the direct dissociation of molecular nitrogen upon electron exposure of the ice and the consequent recombination of an electronically excited (and/or suprathermal) nitrogen atom with molecular nitrogen [$\textrm{N}_{2}(\widetilde{X}^{1}\Sigma^{+}_{g})+\textrm{N}(^{2}D/^{4}S)\rightarrow{\textrm{N}_{3}(\widetilde{X}^{2}\Pi}$)] has been extensively investigated by different groups by means of energetic processing of N$_2$-rich ices \citep[e.g.,][]{Hudson_Moore2002, Moore_Hudson2003, Baratta_etal2003, Jamieson_Kaiser2007, Wu_etal2012, Wu_etal2013a, Wu_etal2013b, CruzDiaz_etal2014b, Lo_etal2015, Mencos_etal2017, Hudson_Gerakines2019, Lo_etal2019}. Unlike \cite{Wu_etal2013a, Wu_etal2013b}, in our experiments, neither N$^{+}_{2}$ nor N$^{+}_{3}$ are detected after irradiation. Moreover, the intensity increase of the 117\,nm band is apparently not linked to the formation of N$_3$ or N$^{+}_{3}$ because it is clearly visible at low doses, whereas N$_3$ is only detectable at higher doses. Finally, our corresponding infrared data did not show any clear sign of formation products from irradiation of N$_2$. Clearly, the newly formed N$_3$ was below the detection limit of our FT-IR system.

\subsection{O$_2$:N$_2$~=~1:1 mixed ice}
\begin{figure}
\centering
\includegraphics[width=0.5\textwidth]{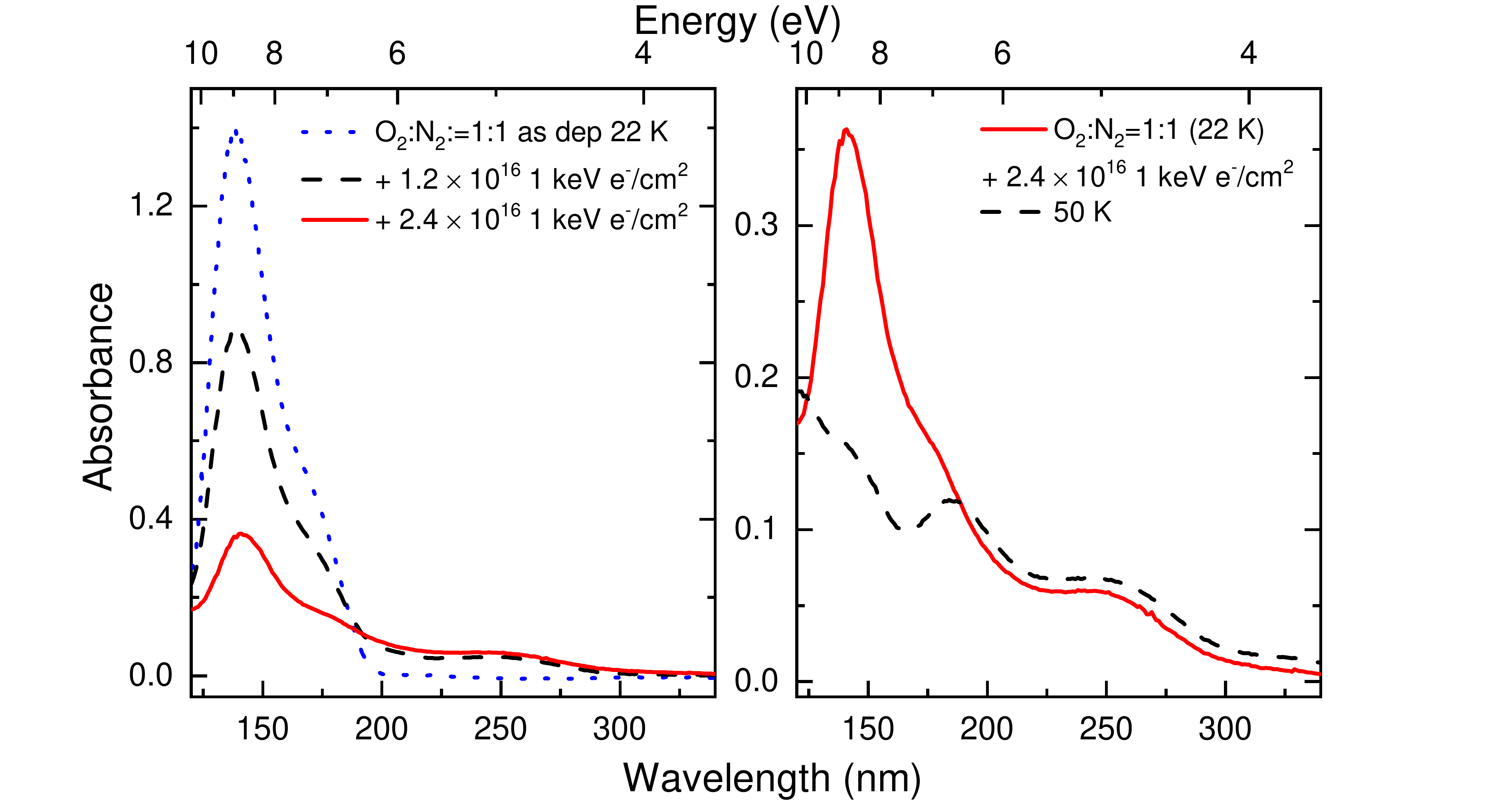}
\vspace{-5mm}
\caption{Left panel: VUV absorbance spectrum ($120-340$~nm) of a frozen mixture O$_2$:N$_2$~=~1:1 as deposited at 22~K (dotted blue line) and after 1\,keV electron irradiation at two different fluences (dashed black and solid red lines). Right panel: The irradiated mixture is shown again (solid red line) on a different scale, along with the spectra obtained after annealing the ice to 50~K (dashed black line).}
\label{N2_O2}
\end{figure}

The VUV absorbance spectrum of a 0.24~\mc\ thick frozen O$_2$:N$_2$~=~1:1 mixture deposited at 22~K is shown in the left panel of Figure~\ref{N2_O2} (dotted blue line) along with those obtained after 1\,keV electron irradiation at two different fluences (1.2 and 2.4~$\times$10$^{16}$\,e$^-$/cm$^{2}$, dashed black and solid red lines, respectively). The average stopping power of a 1\,keV electron impinging a mixture of O$_2$:N$_2$~=~1:1 with an estimated density of 1.24~g/cm$^{3}$ is 4.34$\times~10^{-15}$~eV~cm$^{2}$/16u, while a total fluence of 2.4~$\times$10$^{16}$~e$^-$/cm$^{2}$ corresponds to a dose of 104.3~eV/16u. As mentioned before, because of the different VUV photoabsorption cross sections for O$_2$ and N$_2$, the VUV spectrum of the deposited mixture exhibits a typical profile characteristic of pure O$_2$ ice (see Figure~\ref{O2}), that is, nitrogen molecules are almost not detectable in the mixture. Upon electron bombardment, the spectrum profile changes in a similar way as seen for pure O$_2$ ice, that is, the molecular oxygen bands decrease, while ozone ice is detected at 257~nm. Major differences between the 1\,keV electron irradiated pure O$_2$ and the O$_2$:N$_2$~=~1:1 mixture appear when the ice is warmed up to 50~K. At 50~K, O$_2$ and N$_2$ are expected to be fully desorbed, leaving the most refractory reaction products in the ice. The right panel of Figure~\ref{O2} shows a spectrum that exhibits the features of ozone at 50~K (solid magenta line, see 120, 140, and 257~nm bands), while the corresponding panel in Figure~\ref{N2_O2} shows a spectrum of all the formation products of the reaction \textrm{O}$_{2}$:\textrm{N}$_2+\textrm{e}^-\rightarrow{products}$ at 50~K (dashed black line, see 120, 140, 185, and 257~nm bands).

\begin{figure}
\centering
\includegraphics[width=0.5\textwidth]{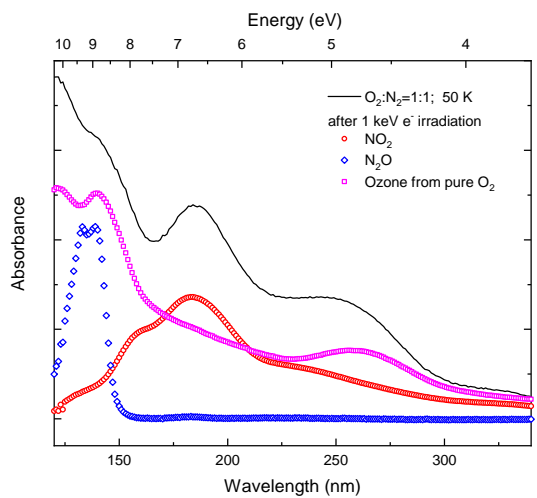}
\caption{VUV absorbance spectrum ($120-340$~nm) of a frozen mixture O$_2$:N$_2$~=~1:1 obtained after electron bombardment and annealing to 50~K (solid black line) is compared to VUV spectra of O$_3$ at 50~K formed after electron irradiation of pure O$_2$ at 22~K (magenta squares), pure deposited NO$_{2}$ (red circles), and N$_2$O ices at 22~K (blue diamonds).}
\label{N2_O2vsN_oxides}
\end{figure}

Laboratory work shows that N$_x$O$_y$ species form in the ice upon UV photon, electron, and ion irradiation of N- and O-rich ices \citep[e.g.,][]{Jamieson_etal2005, Sicilia_etal2012, Boduch_etal2012, Vasconcelos_etal2017, Hudson_2018, Carrascosa_etal2019}. Further energetic and `nonenergetic', that is, thermal surface atom addition reactions, processing of simple nitrogen oxides (e.g., NO, N$_2$O, and NO$_2$) leads to the formation of other nitrogen oxides, as discussed elsewhere \citep[e.g.,][]{Minissale_etal2014, Ioppolo_etal2014, Almeida_etal2017}. To aid in the identification of nitrogen oxides in the laboratory VUV spectrum of the irradiated and annealed O$_{2}$:N$_2$ mixture, we have acquired VUV photoabsorption spectra of pure NO, N$_2$O, and NO$_{2}$ deposited at 22~K. Figure~\ref{N2_O2vsN_oxides} compares the profile of the spectrum of the O$_{2}$:N$_2$ mixture irradiated at 22~K and further warmed up to 50~K with spectra of ozone formed after irradiation of pure O$_2$ and annealed to 50~K (magenta squares), pure N$_2$O at 22~K (blue diamonds), and pure NO$_{2}$ at 22~K (red circles). The peak positions of bands from O$_3$, N$_2$O, and NO$_{2}$ are all in good agreement with those observed in the processed O$_{2}$:N$_2$ mixture. NO is not included in Figure~\ref{N2_O2vsN_oxides} because even if it were formed at 22~K upon 1\,keV electron irradiation of the O$_{2}$:N$_2$ mixture, it would have desorbed at 50~K, that is, during annealing of the ice \citep{Minissale_etal2014}. The comparison of VUV spectral components with the VUV spectrum of an O$_2$:N$_2$ mixture irradiated at 22~K and annealed to 50~K confirms that O$_3$, N$_2$O, and NO$_{2}$ are among the main electron irradiation products in the ice. Moreover, although O$_3$, N$_2$O, and NO$_{2}$ bands present broad profiles in the VUV spectral range, their peaks do not overlap. Therefore the identification of similar slopes in astronomical observations of Solar System icy objects can potentially trace both nitrogen oxides and ozone in such ices.

\subsection{Pure NO, N$_{2}$O, and NO$_{2}$ ices}
\begin{figure}
\centering
\includegraphics[width=0.5\textwidth]{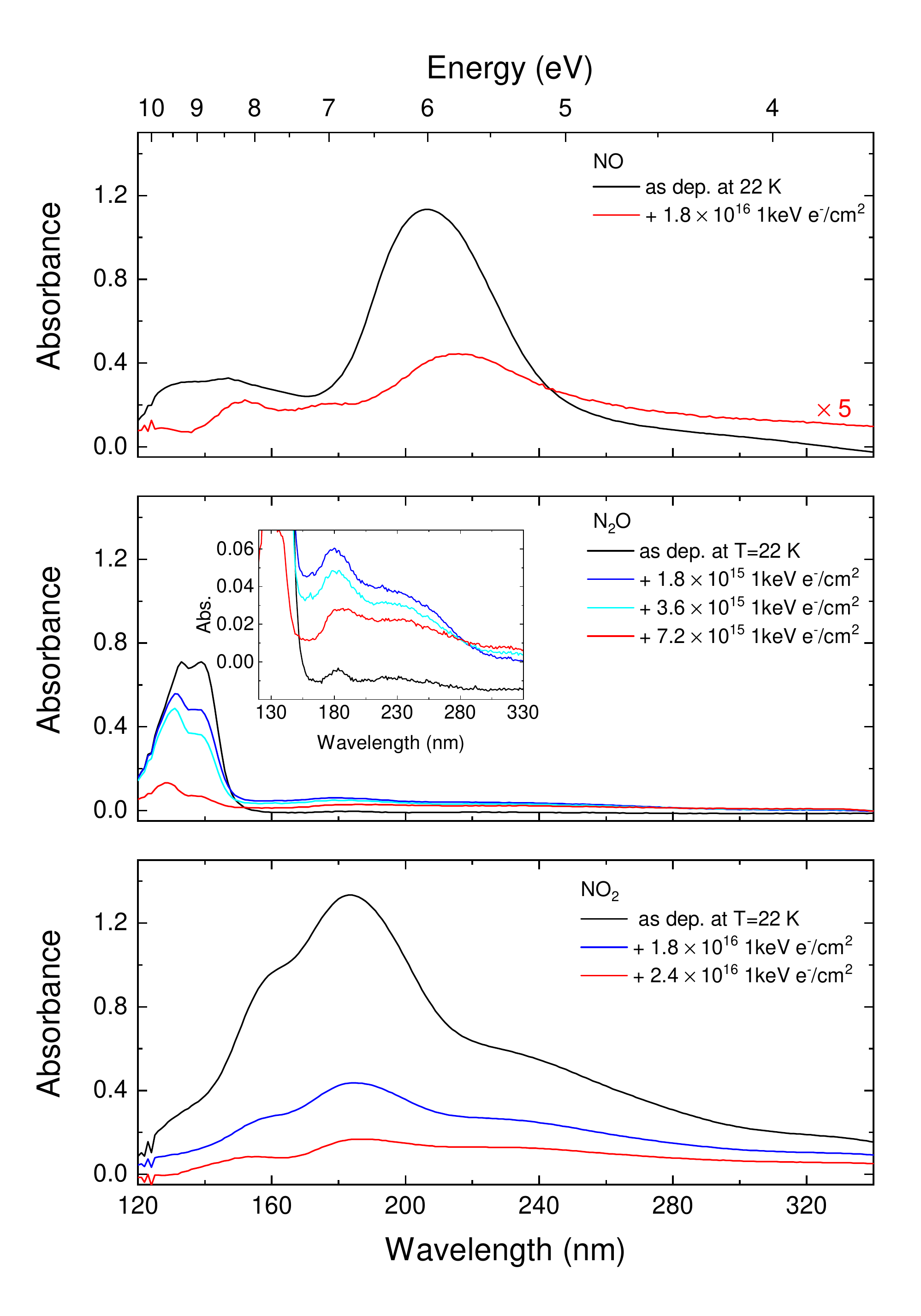}
\caption{VUV absorbance spectra ($120-340$~nm) of pure NO (top panel), N$_2$O (middle panel), and NO$_2$ (bottom panel) ices as deposited at 22~K (solid black lines) and after exposure to 1\,keV electrons.}
\label{N_oxides1}
\end{figure}

To the best of our knowledge, \cite{Lu_etal2008} published the only available VUV absorption spectra of NO and N$_2$O ice deposited at 10~K in the spectral range $110-250$~nm. There are no data in the literature of VUV spectra of NO$_2$ in the condensed phase. We present the first systematic VUV photoabsorption spectra of pure and mixed nitrogen oxides before and after 1\,keV electron exposure in the spectral range $120-340$~nm. Figure~\ref{N_oxides1} shows VUV photoabsorption spectra of pure deposited NO (top panel), N$_{2}$O (mid-panel), and NO$_{2}$ (bottom panel) ices at 22~K. \cite{Lu_etal2008} identified two VUV absorption broad features at 149 and 208~nm for solid NO, with another potential band below 100~nm. They also presented four broad features at 112, 128, 139, and 178~nm for N$_{2}$O ice. Table~\ref{new VUV_peaks1} lists all the absorption features identified in our VUV photoabsorption spectra of pure deposited nitrogen oxides. Our VUV spectrum of pure NO deposited at 22~K does not suggest any indication for a broad feature below 100~nm, while it highlights several broad bands at 136 \citep[not observed by ][]{Lu_etal2008}, 147, and 206~nm, with the latter being also the strongest in intensity by a factor of $\sim$3 with respect to the others. The VUV absorption cross sections of NO molecules in the gas phase consist of band structure superposed on ionization and dissociation continua in the range $50-235$~nm, with the first ionization limit at 135~nm corresponding to the band observed in our VUV spectra of solid NO \citep{Hudson1971}. The slope observed by \cite{Lu_etal2008} at low wavelengths could be due to Rayleigh scattering off islands of material, that is, a rough ice surface. The VUV absorption spectrum of solid N$_{2}$O deposited at 22~K presents two strong broad absorption peaks at 133 and 139~nm and two bands at 183 and 225~nm that are $\sim$2 orders of magnitude weaker than those at lower wavelength (see Figure~\ref{N_oxides1}). \cite{Lu_etal2008} did not identify the 225~nm band due to its weakness in intensity and the fact that it extends beyond their measured spectral range, that is, between $200-300$~nm. Finally, the VUV absorption spectrum of solid NO$_{2}$ deposited at 22~K shows a band shoulder at 132~nm and several other broad intense absorption bands at 163, 183, and 230~nm, with the 183~nm feature being the most intense.

\begin{table}
\centering
        \caption{VUV photoabsorption features detected after deposition of pure nitrogen oxide ices at 22~K and compared to band peak positions of a VUV spectrum of 1\,keV irradiated O$_2$:N$_2$~=~1:1 mixture heated to 50~K. }
        \label{new VUV_peaks1}
        \begin{tabular}{lll}
                \hline
                &\multicolumn{2}{c}{Position (nm)}\\
        \cline{2-3}
        Molecule & This Work$^{a}$ & Literature$^{a,b}$\\
        \hline
        NO              &---       & $<$100 s, sh\\
                        &136 m     & ---         \\
                        &147 m     & 149 m       \\
                        &206 vs    & 208 vs      \\
        N$_2$O          &---       & 112 s       \\
                        &133 vs    & 128 vs      \\
                        &139 vs    & 139 m       \\
                        &183 vw    & 178 vw      \\
                        &225 vw    & ---         \\
       NO$_2$           &132 m, sh &             \\
                        &163 s     &             \\
                        &183 vs    &             \\
                        &230 m, sh &             \\
       O$_2$:N$_2$~=~1:1$^{c}$ &137 vs, sh &     \\
                               &185 s      &     \\
                               &242 m, sh  &     \\
                \hline
        \end{tabular}
\begin{flushleft}
\footnotetext{}{$^{a}$vw = very weak, w = weak, m = medium, s = strong, vs = very strong, sh = shoulder; $^{b}$\cite{Lu_etal2008}; $^{c}$Ice mixture exposed to 1\,keV electrons at 22~K and then heated to 50~K.}
\end{flushleft}
\end{table}

The top panel of Figure~\ref{N_oxides1} shows a VUV photoabsorption spectrum of pure NO (red line) exposed to 1\,keV electron irradiation at 22~K. After a 1\,keV electron dose of 80.2~eV/16u, the strongest NO absorption peak at 206~nm decreases in intensity by a factor of nearly 13 and shifts to 216~nm. The VUV photoabsorption feature at 136~nm disappeared and the 147~nm band shifted to 152~nm. However, it is also possible that the VUV band at 147~nm disappeared as well and that the new 152~nm feature is due to molecular oxygen formed in the ice. Pure molecular oxygen has its strongest band peak position at 154~nm (see Figure~\ref{O2}). However, it should be noted that the 154~nm band of pure O$_2$ is wider than the new 152~nm band observed after irradiation of NO ice. Hence the identification of molecular oxygen is only tentative. Moreover, the large wavelength step size (1~nm) used during the acquisition of all the VUV spectra of nitrogen oxides recorded in this study does not allow for the unambiguous identification of newly formed N$_2$ in the ice. Hence it is not clear whether in the range $115-150$~nm there are any weak features superimposed on the continuum that could be associated with the Lyman-Birge-Hopfield and Tanaka absorption systems of solid N$_{2}$. A VUV spectrum of pure N$_2$ ice acquired with the same step size used for nitrogen oxides reveals that only the broader features in the range $130-145$~nm are detectable in this configuration. We should also point out that N$_{2}$ has a very low VUV absorption cross section compared to O$_{2}$ and other N$_x$O$_y$ species, complicating its identification in the ice. Finally, the VUV spectrum of processed NO presents a new feature at 179~nm that can be attributed to the strongest band of pure NO$_{2}$ molecules (see Table~\ref{new VUV_peaks1}). Although the detection of new species should be considered only tentative, their presence in the ice is expected when their chemical formation pathway is considered. Nitric oxide is indeed dissociated by 1\,keV electrons into atomic nitrogen and oxygen that can then recombine reforming NO or react with the surrounding NO molecules or with other atoms to form new species. It is likely that atoms formed upon electron exposure of the ice are electronically excited and/or suprathermal. Therefore directional diffusion should occur and a direct recombination to reform NO is less likely than the formation of other species. This can also explain the fast decrease of NO as a function of electron dose. O$_{2}$ and N$_{2}$ form through the recombination of O and N atoms, respectively. Nitrogen dioxide is formed through the reaction $\textrm{NO}+\textrm{O}\rightarrow\textrm{NO}_2$. Similarly, nitrous oxide should form through the reaction $\textrm{NO}+\textrm{N}\rightarrow\textrm{N}_2\textrm{O}$. However, as discussed below, N$_2$O is destroyed upon electron exposure at very low doses compared to other N$_x$O$_y$ molecules. Therefore it is likely that, if formed, N$_2$O is already entirely consumed in the NO experiment at a total dose of 80.2~eV/16u, explaining its nondetection in the VUV spectrum of irradiated NO ice.

The middle panel of Figure~\ref{N_oxides1} presents three VUV spectra of electron irradiation of solid N$_2$O at 22~K with a total dose of 27.6~eV/16u. At such low doses compared to other experiments shown in this work, N$_2$O is already nearly entirely destroyed, that is, the VUV band at 133~nm decreases by a factor of 5. Upon electron exposure, the two main band peaks of N$_2$O gradually shift to 129 and 137~nm, respectively. Unfortunately, for the aforementioned reasons it is not clear whether any weak feature superimposed on the continuum are due to solid N$_2$. However, N$_2$ is expected to be formed in the ice because it forms when 1\,keV electrons remove an oxygen from N$_2$O. It is interesting to note that two bands at 181 and 230~nm associated with NO$_2$ ice appear at lower doses and then decrease at higher electron exposures. To form solid NO$_2$, 1\,keV electrons need to break the N$_2$O molecules from both the N and O sides, that is, $\textrm{N}_{2}\textrm{O}+\textrm{e}^-\rightarrow\textrm{N}+\textrm{NO}$ and $\textrm{N}_{2}\textrm{O}+\textrm{e}^-\rightarrow\textrm{O}+\textrm{N}_2$, respectively. The further oxidation of NO leads to the formation of NO$_2$. The detection of NO$_2$ also suggests the presence of N$_2$, O$_2$ , and NO in the ice.

In the bottom panel of Figure~\ref{N_oxides1}, VUV photoabsorption spectra of frozen NO$_2$ exposed to 1\,keV electrons are displayed. Solid NO$_2$ is perhaps the most stable species among the nitrogen oxides studied here when exposed to electron irradiation. Under the same electron doses discussed for solid NO (80.2~eV/16u), the main VUV absorption band of NO$_2$ decreases only by a factor of 3. Upon irradiation, the absorption feature at 132 nm disappears, while the 163, 183, and 230~nm bands shift to 153, 187, and 226~nm, respectively. As for the case of NO, the band at 153~nm could potentially be, at least in part, due to O$_2$ molecules because its shape changes more than other bands in the same spectrum. Although NO is expected to be the most likely first species formed after electron irradiation of frozen NO$_2$ ($\textrm{NO}_{2}+\textrm{e}^-\rightarrow\textrm{O}+\textrm{NO}$), its detection is not possible in the VUV spectra of processed ice. One possible reason is the broad nature and intensity of the VUV NO$_2$ features, which cover almost the full $120-300$~nm spectral range.

\subsection{Binary ice mixtures containing NO, N$_{2}$O, and NO$_{2}$}
\begin{figure}
\centering
\includegraphics[width=0.5\textwidth]{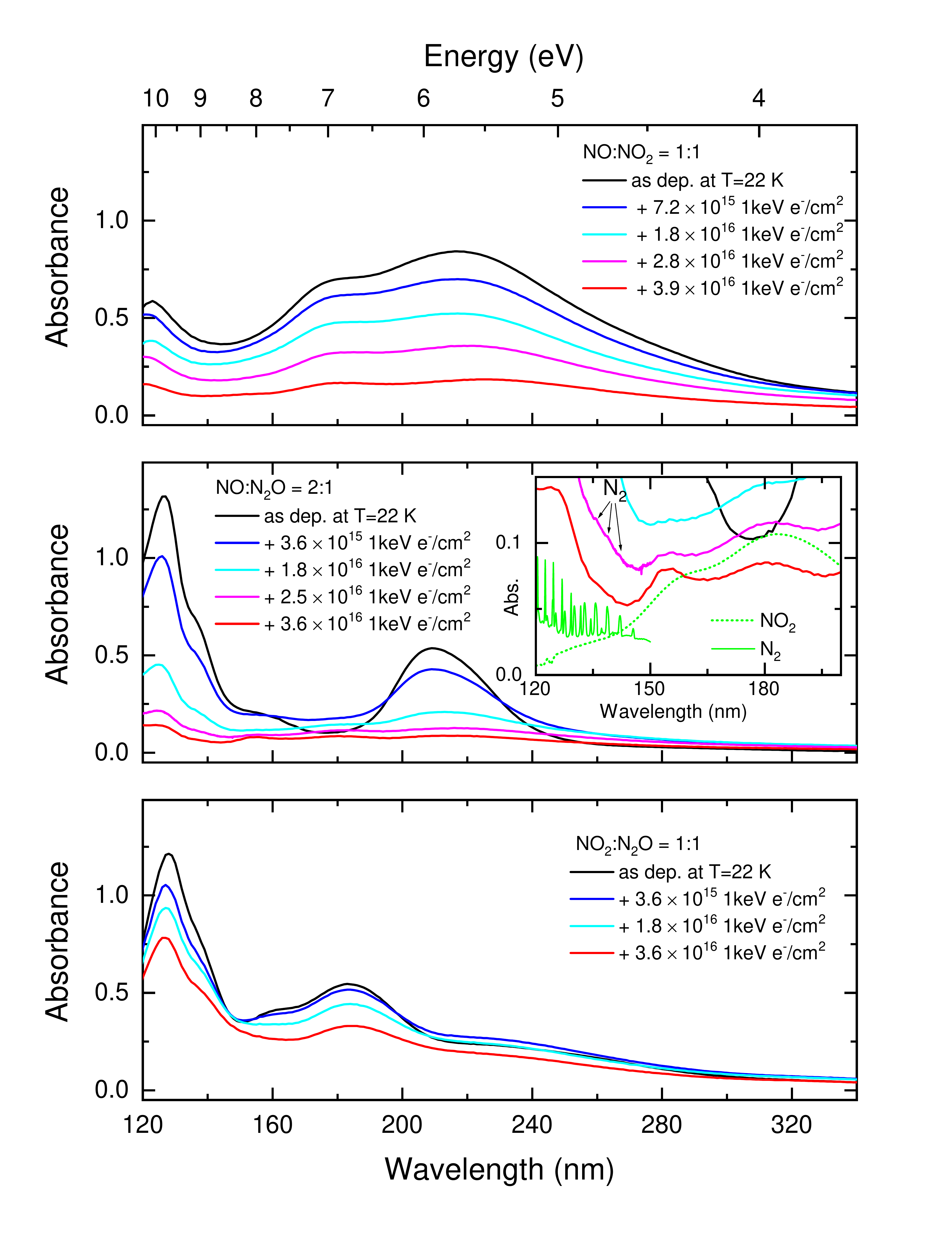}
\caption{VUV absorbance spectra ($120-340$~nm) of NO:NO$_2$~=~1:1 (top panel), NO:N$_2$O~=~2:1 (mid-panel), and NO$_2$:N$_2$O~=~1:1 (bottom panel) ice mixtures as deposited at 22~K (solid black lines) and after exposure to 1\,keV electrons.}
\label{N_oxides2}
\end{figure}

The VUV photoabsorption spectrum of the NO:NO$_2$~=~1:1 mixture deposited at 22~K shows a unique profile with absorption bands at 123, 184, and 217~nm (see the top panel of Figure~\ref{N_oxides2} and Table~\ref{new VUV_peaks2}). The mathematical addition of pure NO and NO$_2$ VUV spectra leads to a completely different spectral profile with peaks at 133 (weak), 160 (medium), and 199~nm (very strong), not shown in Figure~\ref{N_oxides2}. The relative intensity and peak position differences between the mathematically added spectrum of the pure components and the experimentally measured ice mixture indicate that there is a strong molecular interaction between the two ice components in the mixture. This is further confirmed by the fact that 1\,keV electron irradiation of the ice mixture causes the destruction of the ice, that is, decrease of absorption band intensities, without the appearance of any other clear absorption band due to newly formed species. However, we should point out that the nondetection of newly formed species can be due to the broad nature of the absorption bands of the investigated mixture that can cover signals from other absorption species.

\begin{table}
\centering
        \caption{VUV photoabsorption features detected after deposition of nitrogen oxide ice mixtures at 22~K. }
        \label{new VUV_peaks2}
        \begin{tabular}{ll}
                \hline
        Molecule & Position (nm)$^{a}$\\
        \hline
        NO:NO$_2$~=~1:1         & 123 s\\
                                & 184 s, sh\\
                                & 217 vs\\
        NO:N$_2$O~=~2:1         & 126 vs\\
                                & 135 s, sh\\
                                & 154 m, sh\\
                                & 209 s\\
        NO$_2$:N$_2$O~=~1:1     & 128 vs\\
                                & 136 s, sh\\
                                & 163 m, sh\\
                                & 183 s\\
                                & 222 w, sh\\
                \hline
        \end{tabular}
\begin{flushleft}
\footnotetext{}{$^{a}$w = weak, m = medium, s = strong, vs = very strong, sh = shoulder.}
\end{flushleft}
\end{table}

The middle panel of Figure~\ref{N_oxides2} shows VUV photoabsorption spectra of a mixture NO:N$_2$O~=~2:1 at 22~K before and after 1\,keV electron irradiation. The choice of depositing a 2:1 mixture is due to the higher absorption cross section of solid N$_2$O compared to NO ice when the two components are mixed. Table~\ref{new VUV_peaks2} lists all the absorption bands visible upon deposition. There is little shift of all bands compared to the respective ones of pure ice components. However, the N$_2$O band at 126~nm is now much stronger than the other band at 135~nm. This is the case when solid N$_2$O is also mixed with other nitrogen oxides. In pure N$_2$O ice the corresponding bands at 133 and 139~nm are of equal strength, see Figure~\ref{N_oxides1} and Table~\ref{new VUV_peaks1}. Interestingly, the wide 154~nm band disappears at a fluence higher than 1~$\times$10$^{16}$\,e$^-$/cm$^{2}$ and a new narrower feature appears at the same wavelength, resembling the band observed upon electron irradiation of pure NO. As for pure NO, we tentatively assigned this band to pure oxygen. The inset of the middle panel of Figure~\ref{N_oxides2} shows the specific spectral region where several bands of pure N$_2$ at 136, 139, and 142~nm, and pure NO$_2$ at 181~nm are visible. Solid NO$_2$ was detected in the irradiation experiments of pure NO and N$_2$O ices. Therefore its detection is expected. However, this is the only case where absorption features of N$_2$ ice are visible in the VUV spectra of irradiated nitrogen oxide ices. We would like to mention that identification of newly formed N$_2$ ice in the MIR is in general not possible. Hence VUV photoabsorption data of irradiated nitrogen oxide ices can provide further complementary information on the surface reaction network at play.

The VUV photoabsorption spectrum of the NO$_2$:N$_2$O~=~1:1 mixture deposited at 22~K presents absorptions that are similar in peak position and band shape to those due to N$_2$O mixed with NO and those of pure NO$_2$ ice. Upon irradiation, the 163~nm band decreases more rapidly than the rest and the 222~nm band increases first before decreasing as the ice likely becomes thinner as a function of the dose. No new molecules are clearly seen during this experiment. However, it is worth noting that each single ice component is more resistent to electron irradiation when deposited in an ice mixture. Particularly, N$_2$O is still largely present in the ice at all investigated doses, which exceed those used for the irradiation of pure N$_2$O ice.

\section{Astrophysical Implications}

\subsection{Energetic processing of ices in space}

Most of the icy moons belonging to the giant planets of the Solar System are embedded in the magnetospheres
of their respective planets and consequently are exposed to a very intense bombardment by energetic ions (e.g., 20~keV~$-$~100~MeV in the Jupiter magnetosphere) and electrons (e.g., $20-700$~keV in the Jupiter magnetosphere). For instance, the Jupiter moons are embedded in the magnetosphere of the gas giant, and their surfaces are continuously bombarded by energetic ions (mainly H$^{+}$, S$^{n+}$, and O$^{n+}$) accelerated by the magnetic field of Jupiter. As a consequence, the mean energy flux (keV~cm$^{-2}$s$^{-1}$) for Europa, Ganymede, and Callisto are estimated to be $8\times10^{10}$, $5\times10^9$, and $2\times10^8$, respectively \citep{Cooper_etal2001}. As pointed out by \cite{Cooper_etal2001}, the assumed ages (i.e., the time necessary for a complete resurfacing) of the Jovian satellite surfaces are between $10^7-10^9$~years, therefore all the irradiation fluences selected in our work are well within the mean energy fluences that surfaces of Jupiter satellites are exposed to during their lifetimes.

In addition, the surfaces of objects beyond Neptune, which includes the several families of trans-Neptunian objects (TNOs) and KBOs such as Pluto, and comets in the Oort cloud are exposed to galactic cosmic rays and solar wind ion bombardment for billions of years, and therefore there is sufficient accumulation of induced physical and chemical effects \citep[i.e., doses between a few and some hundred eV/16u;][]{Strazzulla_etal2003}. These estimates also agree well with our selected laboratory irradiation doses. Finally, in star-forming regions, interstellar ice grains are exposed to X-rays, UV photons, electrons, and cosmic ions, which induce a complex chemistry within ice layers and the ice-grain interface. The lifetimes of pre- and protostellar phases have been estimated to be about $10^{5}-10^{7}$~years \citep{Caselli_Ceccarelli2012}, thus interstellar grains may accumulate doses of about at least a few eV/16u before ice desorption \citep{Kanuchova_etal2016}. In comparison to our experiments, all the maximum doses used in our experimental work are greater than the estimated values shown in \cite{Kanuchova_etal2016}. However, our experiments show that with the exception of N$_3$, reaction products are clearly formed even at lower doses and can be detected in the VUV and MIR spectral range. Therefore, the N- and O-rich ice chemistry discussed here is also relevant to the ISM. Furthermore, it is relevant to say that we assumed here that the effects induced by energetic processing of ices depend on the deposited dose and not on the specific agent (ions versus electrons). This is justified by the fact that most of the effects produced by ions are due to the secondary electrons produced along the ion track in the irradiated ice \citep[e.g.,][]{Baratta_etal2002}.

\begin{figure}
\centering
\includegraphics[width=0.5\textwidth]{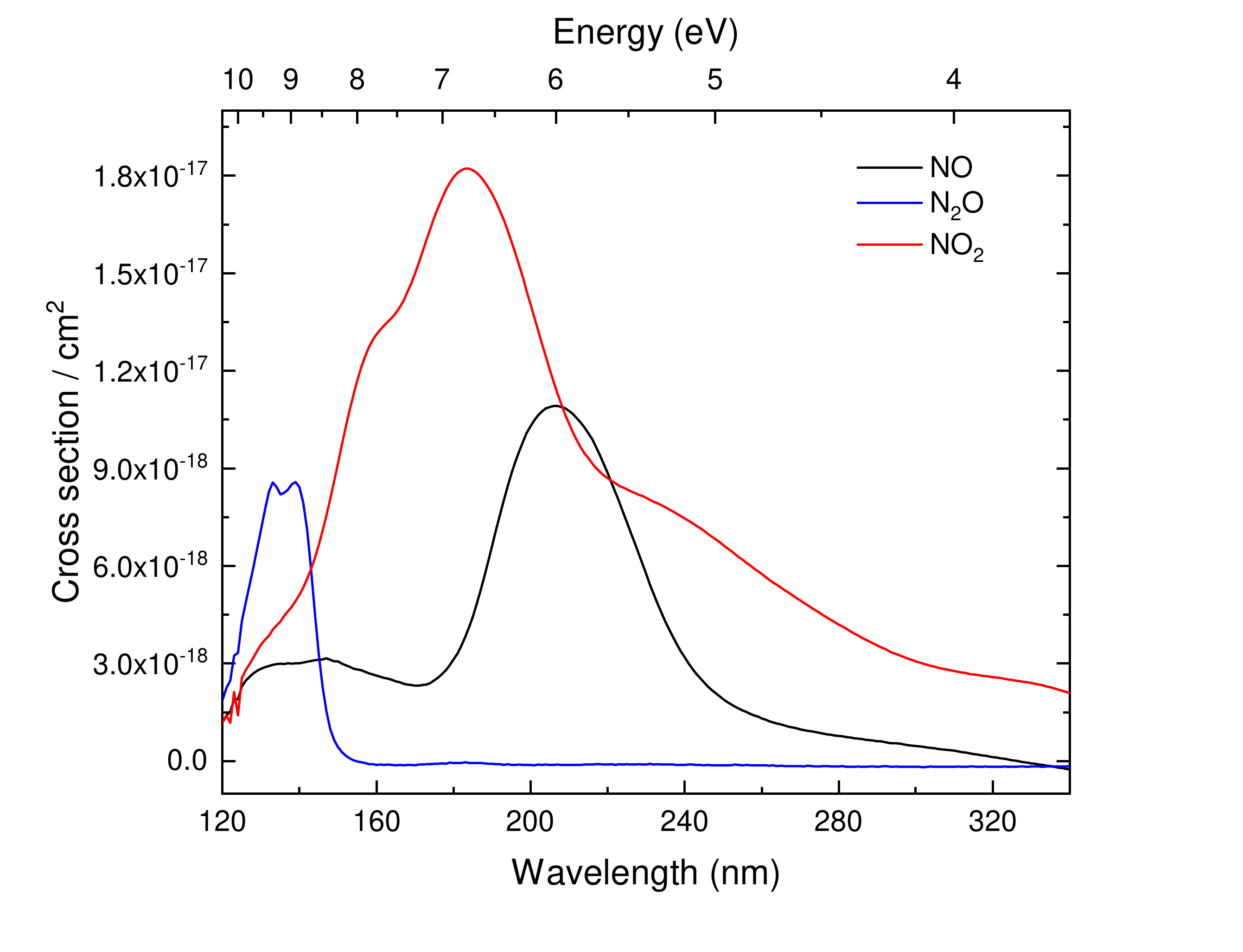}
\caption{VUV photoabsorption cross sections for pure NO, NO$_2$, and N$_2$O ices deposited at 22~K.}
\label{CrossSections}
\end{figure}

\subsection{Photodesorption of nitrogen oxide-containing ices}
The photodesorption rates of small molecules such as O$_{2}$, N$_{2}$, and CO ice measured at different photon energies \citep{Fayolle_etal2011} are clearly correlated to the VUV photoabsorption spectrum of the same material \citep{Lu_etal2008, MunozCaro_etal2016} for the same photon energies. This indicates that photodesorption of some simple ice species is mainly driven by a desorption induced by electronic transition (DIET) process \citep{Fayolle_etal2011, Fayolle_etal2013}. Recently, the photodesorption of solid NO from pure NO ices was found to be as efficient as the photodesorption of CO from pure CO ices, that is, with a yield being around 10$^2$ molecules per incident photon for UV fields relevant to the diffuse and dense ISM \citep{Dupuy_etal2017}. VUV photoabsorption data from \cite{Lu_etal2008} and this work show that the photodesorption of NO follows its VUV photoabsorption spectrum acquired at low temperatures. To the best of our knowledge, photodesorption studies of other pure and mixed nitrogen oxides are not available in the literature. VUV absorption cross-section data obtained in the laboratory allow for a more quantitative study of photon absorption in ice mantles. Figure~\ref{CrossSections} shows the cross-section data obtained from our experimental work on deposited nitrogen oxide ices. Based on our VUV spectroscopic data presented here, the VUV light that reaches an interstellar ice mantle must have an energy higher than 5.5, 6.0, and 8.5~eV to be efficiently absorbed by solid NO, NO$_{2}$, and N$_{2}$O, respectively. Furthermore, we also showed that mixtures containing NO:NO$_2$ present unique features that start to peak around 5 eV and that cannot be reproduced by mathematically adding VUV spectra of pure ice components. This highlights the importance of performing dedicated systematic VUV photoabsorption spectroscopic studies of ices relevant to the Solar System and the ISM in support of future frequency-dependent photodesorption studies of the same ice material.

\subsection{Future observations of nitrogen oxide ices}
In star-forming regions, the exposure of ice grains to energetic processing such as galactic cosmic rays, electrons, UV photons produced by field stars and forming protostars, cosmic-ray-induced UV photons within dark clouds, and thermal heating can all contribute to the formation of larger complex solid species. Here we showed that 1\,keV electron irradiation of an O$_2$:N$_2$ ice leads to the formation of ozone and nitrogen oxides. Further 1\,keV electron irradiation of N$_x$O$_y$ species, pure and in mixtures, causes the formation of simpler and more complex N- and O-bearing species. The energetic and nonenergetic surface chemistry involving nitrogen oxides can lead to the formation of a variety of frozen species, including potential precursors to amino acids such as hydroxylamine \citep[NH$_2$OH;][]{Hudson_2018, Congiu_etal2012, Minissale_etal2014, Ioppolo_etal2014}. Recently, the $Rosetta$ mission highlighted the link between prestellar O$_2$-containing ices and O$_2$-rich cometary ices \citep{Bieler_etal2015}. Hence, understanding the chemical evolution of molecular oxygen and nitrogen in space can potentially shed light on the formation of complex organic species and possibly life-related species throughout the star formation process. Our work adds to the laboratory investigation of O- and N-bearing species exposed to energetic processing by providing highly needed VUV photoabsorption spectra of pure and mixed ices that can be used to understand nonthermal photodesorption of N- and O-bearing molecules from ice grains in cold regions of the ISM as well as protostar photoattenuation effects due to ice material in the midplane of accretion disks in a forming solar-like system. Our results also support the need for a deep search for nitrogen oxides in interstellar ices by means of the MIRI instrument on board the JWST towards different environments in the ISM.

\begin{figure}
\centering
\includegraphics[width=0.5\textwidth]{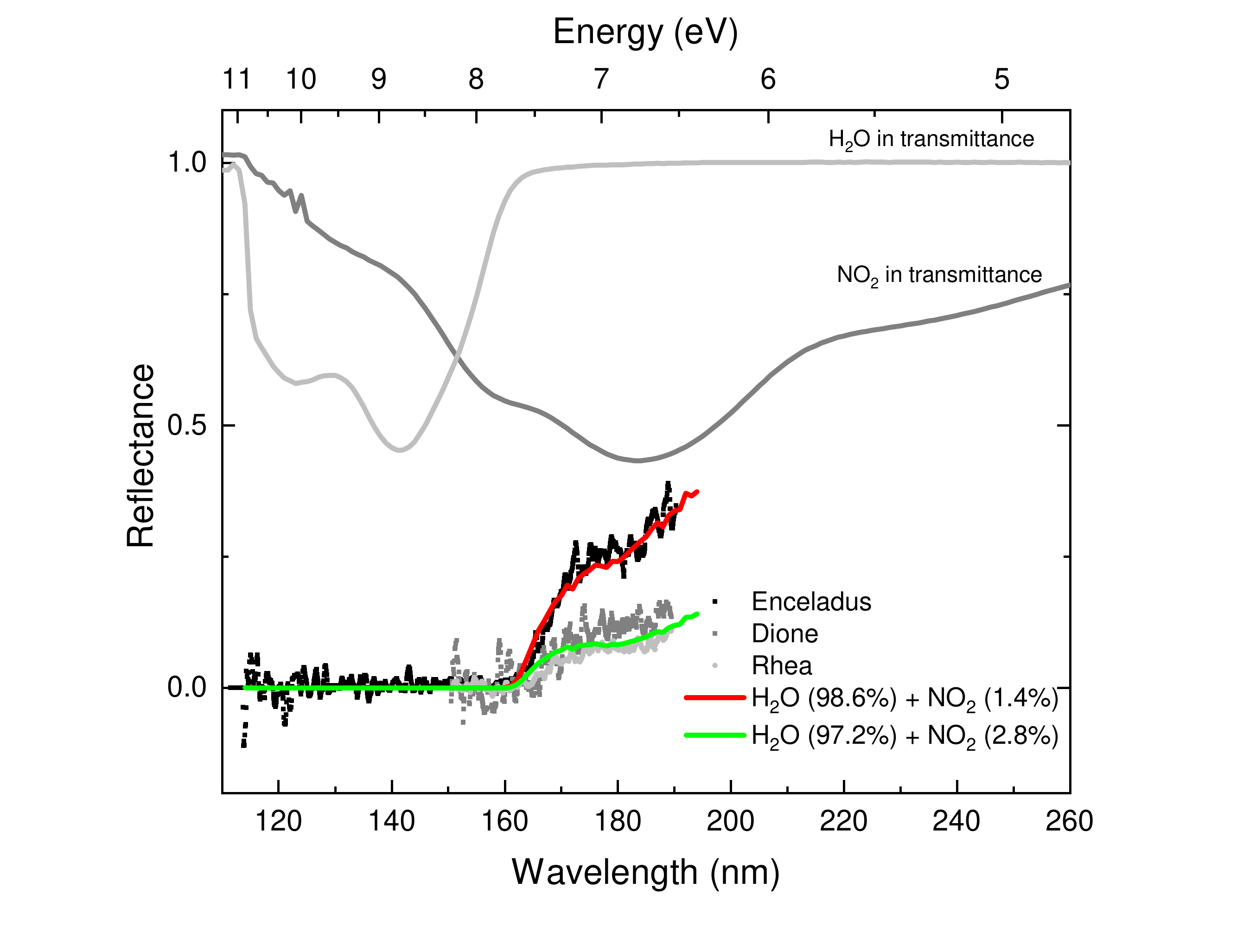}
\caption{VUV ($115-190$~nm) Cassini UVIS data of the surface of Enceladus (square black symbols), Dione (square dark gray symbols), and Rhea (square light gray symbols) compared to a two-fit components (solid red and green lines) of the VUV photoabsorption spectra of solid water and nitrogen dioxide deposited at 22~K (solid light and dark gray lines, respectively). The laboratory ice components are normalized and scaled in the figure for clarity.}
\label{Cassini}
\end{figure}

Our work is relevant to the Solar System as well. The reflectance spectrum of the Saturn moons, such as Enceladus, was recently measured in the VUV ($115-190$~nm) by the Cassini UltraViolet Imaging Spectrograph (UVIS). In the visible and near-infrared, the Enceladus reflectance spectrum is bright, with a surface composed primarily of H$_2$O ice. However, the corresponding VUV spectrum of the surface of the moon is darker than would be expected for pure water ice. \cite{Hendrix_etal2010} explained this surprising finding by the presence of small amounts of NH$_3$ and tholin in addition to H$_2$O ice on the surface. We note that the observational spectrum is fully saturated below 160~nm. Similar VUV spectral profiles were observed on other icy moons of Saturn, for example, Rhea and Dione \citep[e.g., ][]{Royer_Hendrix2014}. In Figure~\ref{Cassini} the VUV reflectance spectra of the ice surfaces of Enceladus, Rhea, and Dione observed by the Cassini spacecraft (square black and gray symbols) are compared to a two-component fit of VUV transmittance spectra of solid amorphous water and NO$_2$ ice. In our fit, frozen nitrogen dioxide is needed to reproduce the slope at 180~nm observed on the surface of the Saturn moons. Our VUV set of data shows that a strong absorption band at 180~nm is visible in the spectra only when NO$_2$ molecules are present in the ice. This is also the case for the irradiated O$_2$:N$_2$ ice mixture. \cite{Hendrix_etal2010} fit the same slope at 180~nm using NH$_3$ ice in combination with tholin and water because the VUV spectrum of ammonia ice presents a band at 180~nm of medium relative intensity, and observations of Enceladus plumes showed NH$_3$ in the lunar ice \citep[$0.8\%$ with respect to water ice, ][]{Waite_etal2006, Waite_etal2009, Pavithraa_etal2018}. However, the same observations suggested molecular nitrogen in the ice \citep[$1\%$ with respect to water ice, ][]{Waite_etal2009}. Therefore, as shown in this work, it is likely that nitrogen oxides are formed through processing of N- and O-bearing ice species on the surface of the Saturn ice moons.

Figure~\ref{Cassini} suggests that $1.4\%$ of the total fit of Enceladus is due to NO$_2$, while double that contribution is needed to fit the slopes of Rhea and Dione. For completeness, we fit the $160-200$~nm slope of Enceladus with other nitrogen- and oxygen-rich ice-fit components. For instance, the addition of a contribution of $>30\%$ of N$_2$O ice, not shown in the figure, reduces the NO$_2$ and water components while still fitting the slope profile well. However, an N$_2$O contribution larger than $30\%$ to the fit is not supported by observational evidence. We also used the VUV spectrum of the O$_2$:N$_2$~=~1:1 ice mixture irradiated and heated to 50~K as a fit component together with water, not shown in the figure, and found that the observed profile of Enceladus is well reproduced by a $<10\%$ contribution of the irradiated ice mixture combined with water ice. We note that our VUV spectrum of O$_2$:N$_2$~=~1:1 irradiated and heated to 50~K presents a broad band due to O$_3$ at 250~nm. No such absorption is visible in observational data of Enceladus \citep{Zastrow_etal2012} and is not expected either due to the desorption temperature of ozone (60~K) and the surface temperature of the moon (72~K). Unfortunately, we do not have a VUV spectrum of the same ice at 70~K. We therefore decided to use the 50~K annealed-ice component. Finally, we stress that the outcome of our fit shown in Figure~\ref{Cassini} does not correspond to the real abundance of the relative species. To carefully quantify the amount of nitrogen oxides in the observed data, we would need optical constants of all materials at the correct temperatures, we should also include other molecules in the fit such as ammonia, and use an appropriate model. Our fit presented in Figure~\ref{Cassini} is performed to show that the simple addition of a small amount of nitrogen dioxide in the fit can reproduce the slope observed on Enceladus, Rhea, and Dione. Hence a two-component fit, including nitrogen oxides, is qualitatively as good as or better than the three-component fit used by \cite{Hendrix_etal2010}. The possible presence of nitrogen oxides on the icy moons of Saturn based on a combined laboratory and observational evidence strongly suggests the need for a deep search of nitrogen oxides on the ice surface of the moon of the giant gaseous planets and the outer Solar System objects.

\section{Conclusions}
We presented new VUV absorption spectra of pure and mixed nitrogen- and oxygen-bearing ices exposed to 1\,keV electron irradiation under conditions relevant to the ISM and the Solar System. Below we list the main findings of this work.

The 1\,keV electron irradiation of solid pure molecular oxygen and nitrogen leads to the formation of frozen ozone and the azide radical, respectively. The analogous irradiation of a O$_2$:N$_2$~=~1:1 ice mixture shows the formation of ozone together with nitrogen oxides, such as N$_2$O, NO$_2$, and most likely NO.

A systematic VUV photoabsorption study of solid nitrogen oxides is presented for the first time. VUV photoabsorption spectra of pure deposited NO, N$_2$O, and NO$_2$ ices were discussed first. VUV spectra of binary mixtures of nitrogen oxides show unique peaks indicating a strong molecular interaction within the mixed material. This cannot be reproduced by simply adding pure ice components to a spectral fit.

Pure and mixed nitrogen oxide ices were further exposed to 1\,keV electron irradiation to study the energetic chemistry induced by electrons in the ices and the durability of such materials as a function of the electron dose. In addition to the formation and destruction of nitrogen oxides, molecular oxygen and nitrogen are also observed in the VUV spectral range upon electron irradiation. The observation of homonuclear species in the VUV spectral range complements literature laboratory data in the MIR.

VUV photoabsorption spectra and cross sections of pure deposited NO, N$_2$O, and NO$_2$ ices can be used to understand nonthermal photodesorption of N- and O-bearing molecules from ice grains in cold regions of the ISM as well as protostar photoattenuation effects due to ice material in the midplane of accretion disks in a forming solar-like system.

We reproduce the reflectance spectral profiles of Enceladus, Rhea, and Dione in the $115-190$~nm range with a binary H$_2$O and NO$_2$ component fit. Irradiated O$_2$:N$_2$ mixtures can also qualitatively reproduce the observed profile, indicating that nitrogen oxides are likely present on the surface of the icy Saturn moons. Previous studies never considered nitrogen oxides as possible candidates for the observed darkening of the reflectance spectra of the icy Saturn moon in the VUV range.

Hence our results support the need for a deep search for nitrogen oxides in interstellar ices in the MIR by means of JWST observations toward different environments in the ISM as well as in the Solar System in the VUV spectral range by the JUICE mission.

\begin{acknowledgements}
The authors thank Dr. Amanda R. Hendrix for providing observational Cassini's UVIS data and an anonymous reviewer whose suggestions helped improve and clarify this manuscript. The research presented in this work has been supported by the Royal Society University Research Fellowship (UF130409), the Royal Society Research Fellow Enhancement Award (RGF/EA/180306), and the project CALIPSOplus under the Grant Agreement 730872 from the EU Framework Programme for Research and Innovation HORIZON 2020. Furthermore, S.I. recognizes the Royal Society for financial support. The research of Z.K. was supported by VEGA – the Slovak Grant Agency for Science (grant No. 2/0023/18) and COST Action TD 1308. R.L.J. acknowledges the STFC for her Ph.D. Studentship under grant no. ST/N50421X/1. A.D. acknowledges the Daphne Jackson Trust and Leverhulme Trust (grant no. ECF/2016-842) for her Fellowships. R.L.J. and A.D. also acknowledge the Open University for financial support. N.J.M. acknowledges support from the Europlanet 2020 RI from the European Union's Horizon 2020 research and innovation programme under grant agreement No 654208. G.S. was supported by the Italian Space Agency (ASI 2013-056 JUICE Partecipazione Italiana alla fase A/B1) and by the European COST Action TD1308-Origins and evolution of life on Earth and in the Universe (ORIGINS).
\end{acknowledgements}


\begin{thebibliography}{}

\bibitem[\protect\citeauthoryear{Almeida et al.}{2017}]{Almeida_etal2017} Almeida, G. C., Pilling, S., de Barros, A. L. F., et al. 2017, MNRAS, 471, 1330
\bibitem[\protect\citeauthoryear{Banks}{2012}]{Banks2012} Banks, M. 2012, Phys. World, 25, 06
\bibitem[\protect\citeauthoryear{Baratta \& Palumbo}{1998}]{Baratta_Palumbo1998} Baratta, G. A., \& Palumbo, M. E. 1998, JOSAA, 15, 3076
\bibitem[\protect\citeauthoryear{Baratta et al.}{2002}]{Baratta_etal2002} Baratta, G. A., Leto, G., \& Palumbo, M. E. 2002, A\&A, 384, 343
\bibitem[\protect\citeauthoryear{Baratta et al.}{2003}]{Baratta_etal2003} Baratta, G. A., Domingo, M., Ferini, G., et al. 2003, NIMPB, 209, 283
\bibitem[\protect\citeauthoryear{Bar-Nun et al.}{2007}]{BarNun_etal2007} Bar-Nun, A., Notesco, G., \& Owen, T. 2007, Icarus, 190, 655
\bibitem[\protect\citeauthoryear{Bennett \& Kaiser}{2005}]{Bennett_Kaiser2005} Bennett, C. J., \& Kaiser, R. I. 2005, ApJ, 635, 1362
\bibitem[\protect\citeauthoryear{Berland et al.}{1994}]{Berland_etal1994} Berland, B. S., Haynes, D. R., Foster, K. L., et al. 1994, JPC, 98, 4358
\bibitem[\protect\citeauthoryear{Bieler et al.}{2015}]{Bieler_etal2015} Bieler, A., Altwegg, K., Balsiger, H., et al. 2015, Nature, 526, 678
\bibitem[\protect\citeauthoryear{Boduch et al.}{2012}]{Boduch_etal2012} Boduch, P., Domaracka, A., Fulvio, D., et al. 2012, A\&A, 544, A30
\bibitem[\protect\citeauthoryear{Boduch et al.}{2016}]{Boduch_etal2016} Boduch, P., Brunetto, R., Ding, J. J., et al. 2016, Icarus, 277, 424
\bibitem[\protect\citeauthoryear{Born \& Wolf}{1970}]{Born_Wolf1970} Born, M., \& Wolf, E. 1970, Pergamon Press, 4th ed.
\bibitem[\protect\citeauthoryear{Bouwman et al.}{2008}]{Bouwman_etal2009} Bouwman, J., Paardekooper, D. M., Cuppen, H. M., Linnartz, H., \& Allamandola, L. J. 2009, ApJ, 700, 56
\bibitem[\protect\citeauthoryear{Boogert et al.}{2015}]{Boogert_etal2015} Boogert, A.C. A., Gerakines, P. A., \& Whittet, D. C. B. 2015, ARA\&A, 53, 541
\bibitem[\protect\citeauthoryear{Brosset et al.}{1993}]{Brosset_etal1993} Brosset, P., Dahoo, R., Gauthierroy, B., Abouafmarguin, L., \& Lakhlifi, A. 1993, Chem. Phys., 172, 315.
\bibitem[\protect\citeauthoryear{Carrascosa et al.}{2019}]{Carrascosa_etal2019} Carrascosa, H., Hsiao, L.-C., Sie, N.-E., Mu\~{n}oz Caro, G. M., \& Chen, Y.-J. 2019, MNRAS, 486, 1985
\bibitem[\protect\citeauthoryear{Caselli \& Ceccarelli}{2012}]{Caselli_Ceccarelli2012} Caselli, P., \& Ceccarelli, C. 2012, AAR, 20, 56
\bibitem[\protect\citeauthoryear{Caselli et al.}{2002}]{Caselli_etal2002} Caselli, P., Walmsley, C. M., Zucconi, A., et al. 2002, ApJ, 565, 331
\bibitem[\protect\citeauthoryear{Chou et al.}{2020}]{Chou_etal2020} Chou, S.-L., Lin, M.-Y., Huang, T.-P., \& Wu, Y.-J. 2020, JMS, 1209, 127954
\bibitem[\protect\citeauthoryear{Congiu et al.}{2012}]{Congiu_etal2012} Congiu, E., Fedoseev, G., Ioppolo, S., et al. 2012, ApJ, 750, L12
\bibitem[\protect\citeauthoryear{Cooper et al.}{2001}]{Cooper_etal2001} Cooper, J. F., Johnson, R. E., Mauk, B. H., Garrett, H. B., \& Gehrels, N. 2001, Icarus, 149, 133
\bibitem[\protect\citeauthoryear{Cooper et al.}{2008}]{Cooper_etal2008} Cooper, P. D., Moore, M. H., \& Hudson, R. L. 2008, Icarus, 194, 379
\bibitem[\protect\citeauthoryear{Cooper et al.}{2010}]{Cooper_etal2010} Cooper, P. D., Moore, M. H., \& Hudson, R. L. 2010, JGR, 115, E10013
\bibitem[\protect\citeauthoryear{Cruikshank et al.}{1993}]{Cruikshank_etal1993} Cruikshank, D. P., Roush, T. L., Owen, T. C., et al. 1993, Science, 261, 742
\bibitem[\protect\citeauthoryear{Cruikshank et al.}{2015}]{Cruikshank_etal2015} Cruikshank, D. P., Grundy, W. M., DeMeo, F. E., et al. 2015, Icarus, 246, 82
\bibitem[\protect\citeauthoryear{Cruz-Diaz et al.}{2014a}]{CruzDiaz_etal2014a} Cruz-Diaz, G. A., Mu\~{n}oz Caro, G. M., Chen, Y.-J., \& Yih, T.-S. 2014, A\&A, 562, A119
\bibitem[\protect\citeauthoryear{Cruz-Diaz et al.}{2014b}]{CruzDiaz_etal2014b} Cruz-Diaz, G. A., Mu\~{n}oz Caro, G. M., Chen, Y.-J., \& Yih, T.-S. 2014, A\&A, 562, A120
\bibitem[\protect\citeauthoryear{Dalton et al.}{2010}]{Dalton_etal2010} Dalton, J. B., Cruikshank, D., Stephan, K., et al. 2010, SSR, 153, 113
\bibitem[\protect\citeauthoryear{Dawes et al.}{2016}]{Dawes_etal2016} Dawes, A., Mason, N. J., \& Fraser, H. J. 2016, PCCP, 18, 1245
\bibitem[\protect\citeauthoryear{Dawes et al.}{2018}]{Dawes_etal2018} Dawes, A., Pascual, N., Mason, N. J., et al. 2018, PCCP, 20, 15273
\bibitem[\protect\citeauthoryear{de Barros}{2015}]{de_Barros_etal2015} de Barros, A. L. F., da Silveira, E. F., Bergantini, A., Rothard, H., \& Boduch, P. 2015, ApJ, 810, 156
\bibitem[\protect\citeauthoryear{Dodge}{1984}]{Dodge1984} Dodge, M. J. 1984, Appl. Optics, 23, 1980
\bibitem[\protect\citeauthoryear{Douglas \& Jones}{1965}]{Douglas_Jones1965} Douglas, A. E., \& Jones, W. J. 1965, CaJPh, 43, 2216
\bibitem[\protect\citeauthoryear{Dows}{1957}]{Dows1957} Dows, D. A., 1957, JCP, 26, 745
\bibitem[\protect\citeauthoryear{Drouin et al.}{2007}]{Drouin_etal2007} Drouin, D., Couture, A. R., Joly, D., et al. 2007, Scanning, 29, 92
\bibitem[\protect\citeauthoryear{Dulieu et al.}{2017}]{Dulieu_etal2017} Dulieu, F., Minissale, M., \& Bockelee-Morvan, D. 2017, A\&A, 597, A56
\bibitem[\protect\citeauthoryear{Dupuy et al.}{2017}]{Dupuy_etal2017} Dupuy, R., F\'{e}raud, G., Bertin, M., et al. 2017, A\&A, 606, L9
\bibitem[\protect\citeauthoryear{Eden et al.}{2006}]{Eden_etal2006} Eden, S., Lim\~{a}o-Vieira, P., Hoffmann, S. V., \& Mason, N. J. 2006, Chem. Phys., 323, 313
\bibitem[\protect\citeauthoryear{Eistrup \& Walsh}{2019}]{Eistrup_Walsh2019} Eistrup, C., \& Walsh, C. 2019, A\&A, 621, A75
\bibitem[\protect\citeauthoryear{Elsila et al.}{1997}]{Elsila_etal1997} Elsila, J., Allamandola, L. J., \& Sandford, S. A. 1997, ApJ, 479, 818
\bibitem[\protect\citeauthoryear{Fayolle et al.}{2011}]{Fayolle_etal2011} Fayolle, E. C., Bertin, M., Romanzin, C., et al. 2011, ApJL, 739, L36
\bibitem[\protect\citeauthoryear{Fayolle et al.}{2013}]{Fayolle_etal2013} Fayolle, E. C., Bertin, M., Romanzin, C., et al. 2013, A\&A, 556, A122
\bibitem[\protect\citeauthoryear{Fateley et al.}{1959}]{Fateley_etal1959} Fateley, W. G., Bent, H. A., \& Crawford Jr., B. 1959, JCP, 31, 204
\bibitem[\protect\citeauthoryear{Fulvio et al.}{2009}]{Fulvio_etal2009} Fulvio, D., Sivaraman, B., Baratta, G. A.,  Palumbo, M. E., \& Mason, N. J. 2009, Spectrochim. Acta A, 72, 1007
\bibitem[\protect\citeauthoryear{Goodman}{1978}]{Goodman1978} Goodman, A. M. 1978, Appl. Optics, 17, 2779
\bibitem[\protect\citeauthoryear{Grundy et al.}{2016}]{Grundy_etal2016} Grundy, W. M., Binzel, R. P., Buratti, B. J., et al. 2016, Science, 351, aad9189
\bibitem[\protect\citeauthoryear{Hendrix, Hansen and Holsclaw}{2010}]{Hendrix_etal2010} Hendrix, A. R., Hansen, C. J., \& Holsclaw, G. M. 2010, Icarus, 206, 608
\bibitem[\protect\citeauthoryear{Herbst \& van Dishoeck}{2009}]{Herbst_etal2009} Herbst, E., \& van Dishoeck, E. F. 2009, ARA\&A, 47, 427
\bibitem[\protect\citeauthoryear{Hincelin et al.}{2011}]{Hincelin_etal2011} Hincelin, U., Wakelam, V., Hersant, F., et al. 2011, A\&A, 530, A61
\bibitem[\protect\citeauthoryear{Holland \& Maier}{1983}]{Holland_Maier1983} Holland, R. F., \& Maier II, W. B. 1983, JCP, 78, 2928
\bibitem[\protect\citeauthoryear{Hudson}{1971}]{Hudson1971} Hudson, R. D. 1971, Rev. Geophys., 9, 305
\bibitem[\protect\citeauthoryear{Hudson \& Moore}{2002}]{Hudson_Moore2002} Hudson, R. L., \& Moore, M. 2002, ApJ, 568, 1095
\bibitem[\protect\citeauthoryear{Hudson et al.}{2008}]{Hudson_etal2008} Hudson, R. L., Palumbo, M. E., Strazzulla, G., et al. 2008, Univ. of Arizona Press, 507
\bibitem[\protect\citeauthoryear{Hudson}{2018}]{Hudson_2018} Hudson, R. L. 2018, ApJ, 867, 160
\bibitem[\protect\citeauthoryear{Hudson \& Gerakines}{2019}]{Hudson_Gerakines2019} Hudson, R. L., \& Gerakines, P. A. 2019, MNRAS, 485, 861
\bibitem[\protect\citeauthoryear{Ioppolo et al.}{2008}]{Ioppolo_etal2008}Ioppolo, S., Cuppen, H. M., Romanzin, C., van Dishoeck, E. F., \& Linnartz, H. 2008, ApJ, 686, 1474
\bibitem[\protect\citeauthoryear{Ioppolo et al.}{2010}]{Ioppolo_etal2010}Ioppolo, S., Cuppen, H. M., Romanzin, C., van Dishoeck, E. F., \& Linnartz, H. 2010, PCCP, 12, 12065
\bibitem[\protect\citeauthoryear{Ioppolo et al.}{2014}]{Ioppolo_etal2014} Ioppolo, S., Fedoseev, G., Minissale, M., et al. 2014, PCCP, 16, 8270
\bibitem[\protect\citeauthoryear{Jamieson et al.}{2005}]{Jamieson_etal2005} Jamieson, C. S., Bennett, C. J., Mebel, A. M., \& Kaiser, R. I. 2005, ApJ, 624, 436
\bibitem[\protect\citeauthoryear{Jamieson \& Kaiser}{2007}]{Jamieson_Kaiser2007} Jamieson, C. S., \& Kaiser, R. I. 2007, CPL, 440, 98
\bibitem[\protect\citeauthoryear{Jones et al.}{2014a}]{Jones_etal2014a} Jones, B. M., Kaiser, R. I., \& Strazzulla, G. 2014, ApJ, 781, 85
\bibitem[\protect\citeauthoryear{Jones et al.}{2014b}]{Jones_etal2014b} Jones, B. M., Kaiser, R. I., \& Strazzulla, G. 2014, ApJ, 788, 170
\bibitem[\protect\citeauthoryear{Johnson et al.}{2019}]{Johnson_etal2019} Johnson, R. E., Oza, A. V., Leblanc, F., et al. 2019, SSR, 215, 20
\bibitem[\protect\citeauthoryear{Ka\v{n}uchov\'{a} et al.}{2016}]{Kanuchova_etal2016} Ka\v{n}uchov\'{a}, Z., Urso, R. G., Baratta, G. A., et al. 2016, A\&A, 585, A155
\bibitem[\protect\citeauthoryear{Knauth et al.}{2004}]{Knauth_etal2004} Knauth, D. C., Andersson, B.-G., McCandliss, S. R., \& Moos, H. W. 2004, Nature, 429, 636
\bibitem[\protect\citeauthoryear{Licandro et al.}{2006}]{Licandro_etal2006} Licandro, J., Pinilla-Alonso, N., Pedani, M., et al. 2006, A\&A, 445, L35
\bibitem[\protect\citeauthoryear{Lo et al.}{2015}]{Lo_etal2015} Lo, J.-I., Chou, S.-L., Peng, Y.-C., et al. 2015, ApJSS, 221, 20
\bibitem[\protect\citeauthoryear{Lo et al.}{2018}]{Lo_etal2018} Lo, J.-I., Chou, S.-L., Peng, Y.-C., et al. 2018, ApJ, 864, 95
\bibitem[\protect\citeauthoryear{Lo et al.}{2019}]{Lo_etal2019} Lo, J.-I., Chou, S.-L., Peng, Y.-C., Lu, H.-C., \& Cheng, B.-M. 2019, ApJ, 877, 27
\bibitem[\protect\citeauthoryear{Loeffler \& Hudson}{2016}]{Loeffler_Hudson2016} Loeffler, M. J., \& Hudson, R. L. 2016, ApJL, 833, L9
\bibitem[\protect\citeauthoryear{Lu et al.}{2008}]{Lu_etal2008} Lu, H.-C., Chen, H.-K., Cheng, B.-M., \& Ogilvie, J. F. 2008, Spectrochim. Acta A, 71, 1485
\bibitem[\protect\citeauthoryear{Luspay-Kuti et al.}{2018}]{Luspay-Kuti_etal2018} Luspay-Kuti, A., Mousis, O., Lunine, J. I., et al. 2018, SSR, 214, 115
\bibitem[\protect\citeauthoryear{Martin-Domenech et al.}{2016}]{MartinDomenech_etal2016} Martin-Domanech, R., Mu\~{n}oz Caro, G. M., \& Cruz-Diaz, G. A. 2016, A\&A, 589, A107
\bibitem[\protect\citeauthoryear{Mason et al.}{2006}]{Mason_etal2006} Mason, N. J., Dawes, A., Holtom, P. D., et al. 2006, Faraday Discuss., 133, 311
\bibitem[\protect\citeauthoryear{Matar et al.}{2008}]{Matar_etal2008} Matar, E., Congiu, E., Dulieu, F., Momeni, A., \& Lemaire, J. L. 2008, A\&A, 492, L17
\bibitem[\protect\citeauthoryear{McClure}{2019}]{McClure_2019} McClure, M. K. 2019, A\&A, 632, A32
\bibitem[\protect\citeauthoryear{Minissale et al.}{2014}]{Minissale_etal2014}  Minissale, M., Fedoseev, G., Congiu, E., et al. 2014, PCCP, 16, 8257
\bibitem[\protect\citeauthoryear{Mencos et al.}{2017}]{Mencos_etal2017} Mencos, A., Nourry, S., \& Krim, L. 2017, MNRAS, 467, 2150
\bibitem[\protect\citeauthoryear{Miyauchi et al.}{2008}]{Miyauchi_etal2008} Miyauchi, N., Hidaka, H., Chigai, T., et al. 2008, CPL, 456, 27
\bibitem[\protect\citeauthoryear{Moore \& Hudson}{2003}]{Moore_Hudson2003} Moore, M. H., \& Hudson, R. L. 2003, Icarus, 161, 486
\bibitem[\protect\citeauthoryear{Mousis et al.}{2016}]{Mousis_etal2016} Mousis, O., Ronnet, T., Brugger, B., et al. 2016, ApJL, 823, L41
\bibitem[\protect\citeauthoryear{Mu\~{n}oz Caro et al.}{2016}]{MunozCaro_etal2016} Mu\~{n}oz Caro, G. M., Chen, Y.-J., Aparicio, S., et al. 2016, A\&A, 589, A19
\bibitem[\protect\citeauthoryear{Niemann et al.}{2005}]{Niemann_etal2005} Niemann, H. B., Atreya, S. K., Bauer, S. J., et al. 2005, Nature, 438, 779
\bibitem[\protect\citeauthoryear{Nour et al.}{1983}]{Nour_etal1983} Nour, E. M., Chen, L. H., \& Laane, J. 1983, JPC, 87, 1113
\bibitem[\protect\citeauthoryear{Owen et al.}{1993}]{Owen_etal1993} Owen, T. C., Roush, T. L., Cruikshank, D. P., et al. 1993, Science, 261, 745
\bibitem[\protect\citeauthoryear{Palmer et al.}{2015}]{Palmer_etal2015} Palmer, M. H., Ridley, T., Hoffmann, S. V., et al. 2015, JCP, 142, 134302
\bibitem[\protect\citeauthoryear{Palumbo et al.}{2008}]{Palumbo_etal2008} Palumbo, M. E., Baratta, G. A., Fulvio, D., et al. 2008, JPCS, 101, 1012002
\bibitem[\protect\citeauthoryear{Pavithraa et al.}{2018}]{Pavithraa_etal2018} Pavithraaa, S., Lo, J.-I., Rahul, K., et al. 2018, Spectrochim. Acta A, 190, 172
\bibitem[\protect\citeauthoryear{Pineau des Forets et al.}{1990}]{Pineau_etal1990} Pineau des Forets, G., Roueff, E., \& Flower, D. R. 1990, MNRAS, 244, 668
\bibitem[\protect\citeauthoryear{Robert Wu et al.}{2002}]{Robert_Wu_etal2002} Robert Wu, C. Y., Judge, D. L., Cheng, B.-M., et al. 2002, Icarus, 156, 456
\bibitem[\protect\citeauthoryear{Royer \& Hendrix}{2014}]{Royer_Hendrix2014} Royer, E. M., \& Hendrix, A. R. 2014, Icarus, 242, 158
\bibitem[\protect\citeauthoryear{Rubin et al.}{2015}]{Rubin_etal2015} Rubin, M., Altwegg, K., van Dishoeck, E. F., \& Schwehm, G. 2015, ApJL, 815, L11
\bibitem[\protect\citeauthoryear{Samuelson et al.}{1997}]{Samuelson_etal1997} Samuelson, R. E., Nath, N. R., \& Borysow, A. 1997, P\&SS, 45, 959
\bibitem[\protect\citeauthoryear{Satorre et al.}{2008}]{Satorre_etal2008} Satorre, M. A., Domingo, M., Millan, C., et al. 2008, P\&SS, 56, 1748
\bibitem[\protect\citeauthoryear{Schmitt et al.}{2017}]{Schmitt_etal2017} Schmitt, B., Philippe, S., Grundy, W. M., and the New Horizons Science Team 2017, Icarus, 287, 229
\bibitem[\protect\citeauthoryear{Sicilia et al.}{2012}]{Sicilia_etal2012} Sicilia, D., Ioppolo, S., Vindigni, T., Baratta, G. A., \& Palumbo, M. E. 2012, A\&A, 543, A155
\bibitem[\protect\citeauthoryear{Sivaraman et al.}{2007}]{Sivaraman_etal2007} Sivaraman, B., Jamieson, C. S., Mason, N. J., \& Kaiser, R. I. 2007, ApJ, 669, 1414
 \bibitem[\protect\citeauthoryear{Sivaraman et al.}{2011}]{Sivaraman_etal2011} Sivaraman, B., Mebel, A. M., Mason, N. J., Babikov, D., \& Kaiser, R. I. 2011, PCCP, 13, 421
\bibitem[\protect\citeauthoryear{Sivaraman et al.}{2014}]{Sivaraman_etal2014} Sivaraman, B., Nair, B.G., Raja Sekhar, B.N., et al. 2014, CPL, 603, 33
\bibitem[\protect\citeauthoryear{Stirling et al.}{1994}]{Stirling_etal1994} Stirling, A., Papai, I., Mink, J., \& Salahub, D., R. 1994, JCP, 100, 2910
\bibitem[\protect\citeauthoryear{Strazzulla \& Johnson}{1991}]{Strazzulla_Johnson1991} Strazzulla, G., \& Johnson, R. E. 1991, Springer, 243
\bibitem[\protect\citeauthoryear{Strazzulla et al.}{2003}]{Strazzulla_etal2003} Strazzulla, G., Cooper, J. F., Christian, E. R., \& Johnson, R. E. 2003,  C. R. Physique, 4, 791
\bibitem[\protect\citeauthoryear{Strazzulla et al.}{2005}]{Strazzulla_etal2005} Strazzulla, G., Leto, G., Spinella, F., \& Gomis, O. 2005, Astrobiology, 5, 612
\bibitem[\protect\citeauthoryear{Taquet et al.}{2016}]{Taquet_etal2016} Taquet, V., Furuya, K., Walsh, C., \& van Dishoeck, E. F. 2016, MNRAS, 462, S99
\bibitem[\protect\citeauthoryear{Teolis et al.}{2006}]{Teolis_etal2006} Teolis, B. D., Loeffler, M. J., Raut, U., Fama, M.,  \& Baragiola, R. A. 2006, ApJ, 644, L141
\bibitem[\protect\citeauthoryear{Teolis et al.}{2017}]{Teolis_etal2017} Teolis, B. D.,  Wyrick, D. Y., Bouquet, A., Magee, B. A.,  Waite Jr., J. H., 2017, Icarus, 284, 18
\bibitem[\protect\citeauthoryear{Tielens \& Hagen}{1982}]{Tielens_etal1982} Tielens, A. G. G. M., \& Hagen, W. 1982, A\&A, 114, 245
\bibitem[\protect\citeauthoryear{van Dishoeck et al.}{2013}]{van_Dishoeck_etal2013} van Dishoeck, E. F., Herbst, E., \& Neufeld D. A. 2013, Chem. Rev., 113, 9043
\bibitem[\protect\citeauthoryear{Vasconcelos et al.}{2017}]{Vasconcelos_etal2017} Vasconcelos, F. A., Pilling, S., Rocha, W. R. M., Rothard, H., \& Boduch, P. 2017, PCCP, 19, 24154
\bibitem[\protect\citeauthoryear{Waite et al.}{2006}]{Waite_etal2006} Waite Jr., J. H., Combi, M. R., Ip, W.-H., et al. 2006, Science, 311, 1419
\bibitem[\protect\citeauthoryear{Waite et al.}{2009}]{Waite_etal2009} Waite Jr., J. H., Lewis, W. S., \& Magee, B. A. 2009, Nature, 460, 487
\bibitem[\protect\citeauthoryear{Westley et al.}{1998}]{Westley_etal1998} Westley, M. S., Baratta, G. A., \& Baragiola, R. A. 1998, JCP, 108, 3321
\bibitem[\protect\citeauthoryear{Wu et al.}{2012}]{Wu_etal2012} Wu, Y.-J., Robert Wu, C. Y., Chou, S.-L., et al. 2012, ApJ, 746, 175
\bibitem[\protect\citeauthoryear{Wu et al.}{2013a}]{Wu_etal2013a} Wu, Y.-J., Chen, H.-F., Chuang, S.-J., \& Huang, T.-P. 2013, ApJ, 768, 83
\bibitem[\protect\citeauthoryear{Wu et al.}{2013b}]{Wu_etal2013b} Wu, Y.-J., Chen, H.-F., Chuang, S.-J., \& Huang, T.-P. 2013, ApJ, 779, 40
\bibitem[\protect\citeauthoryear{Yildiz et al.}{2013}]{Yildiz_etal2013} Yildiz, U. A., Acharyya, K., Goldsmith, P. F., et al. 2013, A\&A, 558, A58
\bibitem[\protect\citeauthoryear{Zastrow et al.}{2012}]{Zastrow_etal2012} Zastrow, M., Clarke, J. T., Hendrix, A. R., \& Noll, K. S. 2012, Icarus, 220, 29
\end{thebibliography}
\end{document}